\documentclass[fleqn,usenatbib]{mnras}
\usepackage{longtable}
\usepackage{newtxtext,newtxmath}

\usepackage[T1]{fontenc}

\DeclareRobustCommand{\VAN}[3]{#2}
\let\VANthebibliography\thebibliography
\def\thebibliography{\DeclareRobustCommand{\VAN}[3]{##3}\VANthebibliography}

\usepackage{graphicx}	
\usepackage{amsmath}	
\usepackage{pdflscape}

\usepackage[dvipsnames]{xcolor}            
\usepackage{ulem}                          
\title[Gaia EDR3 Young Clusters]{Reconstructing Nearby Young Clusters with Gaia EDR3}

\author[Heyl, Caiazzo, Richer  \&  Miller]{
Jeremy Heyl\thanks{email: heyl@phas.ubca.ca}$^1$,
Ilaria Caiazzo\thanks{email: ilariac@caltech.edu; Sherman Fairchild Fellow}$^2$, Harvey Richer$^1$, David R. Miller$^1$
\\
$^{1}$Department of Physics and Astronomy, University of British Columbia, Vancouver, BC V6T 1Z1, Canada\\
$^{2}$TAPIR, Walter Burke Institute for Theoretical Physics, Mail Code 350-17, Caltech, Pasadena, CA 91125, USA\\
}

\date{Accepted XXX. Received YYY; in original form ZZZ}

\pubyear{2021}

\begin{document}
\label{firstpage}
\pagerange{\pageref{firstpage}--\pageref{lastpage}}
\maketitle

\begin{abstract}
We searched through a seven-million cubic-parsec volume surrounding each of the four nearest young open clusters with ages from 40 to 80~Myr  to identify both the current and past members of the clusters within the Gaia EDR3 dataset. We find over 1,700 current cluster members and over 1,200 candidate escapees. Many of these candidates lie well in front and behind the cluster from our point of view, so formerly they were considered cluster members, but their parallaxes put them more than 10~pc from the centre of the cluster today.  We found two candidate high-mass white dwarfs that may have escaped from the alpha Persei cluster and several candidate main-sequence-white-dwarf binaries associated with the younger clusters,  NGC~2451A, IC~2391 and IC~2602.  All of these objects require spectroscopic confirmation. Using these samples of escapee candidates, we develop and implement a novel technique to determine the ages of these clusters and the Pleiades using kinematics with typical uncertainties of $5-7$~Myr.  For all five clusters, this kinematic age is younger than the age estimated with isochrones but within the uncertainties of the isochrone fitting.  We find for the clusters that travel far from the Galactic plane (the Pleiades, NGC~2451A and IC~2602), the formation of the cluster coincides with when the cluster was in the Galactic plane within a few Myr, supporting these age determinations.
\end{abstract}



\section{Introduction}

Globular clusters and open clusters provide crucial benchmarks of stellar evolution. Both types of clusters harbor stars that were all born approximately the same time and with the same composition; Galactic globular clusters however, because of their advanced age, lack stars more massive than the Sun, so one must rely on open clusters to study more massive stars in a similar manner. Although open clusters are typically much closer than globular clusters, allowing more detailed study of individual stars, they are much less rich, numbering hundreds or thousands of stars rather than hundreds of thousands or millions. Furthermore, open clusters lose stars quickly, due both to their size and to their proximity to the Galactic plane, so relatively rare massive and evolved stars that yield the most stringent probes of stellar evolution are often lacking or lost from the cluster.

In \citet[][Paper I hereafter]{pleiades} we developed a technique to trace the lost stars of a cluster back to their birthplace using the five-dimensional phase space information from the Gaia EDR3 catalogue.  We performed this analysis on the Pleiades and probed its internal consistency.  Here, we expand our purview to all the open clusters within 200~pc of the Sun, younger than 200~Myr, to look for evolved stars that may have left the clusters and also to probe the massloss from these clusters. We describe how we construct the search region for each cluster in \S~\ref{sec:sample}, review our geometric technique for finding candidate escapees in \S~\ref{sec:methods} and the properties of the escapee population of the clusters in \S~\ref{sec:results}.  By comparing the properties of the candidate escapees with the putative birth clusters, we implement a novel kinematic method to determine the age of each of these clusters, independently of theories of stellar evolution.  We find that this new technique typically agrees with the ages obtained from isochrones, but with significantly smaller uncertainties.  We present the details of constructing the dataset and the entire catalogue of cluster members and escapee candidates in appendices.

\section{Sample}
\label{sec:sample}

In Paper I we focused on the nearest of the young open clusters, the Pleiades.  Here, we will look at all of the clusters younger than 200~Myr and within 200~pc of the Sun as listed in Tab.~\ref{tab:sample}.  To reduce the size of the samples for each cluster and treat each cluster on an equal footing, we construct the cluster samples using a pair of cone searches through the Gaia EDR3 catalogue centred on each cluster.  The cones are nestled with the narrower cone reaching deeper behind the cluster and a shallower broader cone.  The goal of the nestled cones is to achieve a complete sphere with a radius of 60~pc surrounding each cluster (this sets the minimal parallax and width of the narrower cone) and complete hemisphere of radius 90~pc on the nearside (this sets the minimal parallax and width of the wider cone).
\begin{table*}
    \centering
    \caption{Open clusters within 200~pc younger than 200~Myr from \citet{2018A&A...616A..10G}. I-Age refers to the age determined from fitting stellar isochrones to the data.  K-Age is a age determination from proper motion of the escapees as outlined in \S~\ref{sec:k-ages}.  The typical uncertainty in the isochrone analysis is 20\%, and 5-7~Myr for the kinematic technique.}
    \begin{tabular}{rrrcrrrcrrrc}
    \hline
    \multicolumn{1}{c}{Name} &
    \multicolumn{1}{c}{RA} &
    \multicolumn{1}{c}{Dec} &
    \multicolumn{1}{c}{Distance} &
    \multicolumn{1}{c}{I-Age} &
    \multicolumn{1}{c}{K-Age} &
    \multicolumn{1}{c}{$[Z/H]$} &
    \multicolumn{1}{c}{$E(B-V)$} &
    \multicolumn{1}{c}{$N$} &
    \multicolumn{1}{c}{$\mu_\alpha$} &
    \multicolumn{1}{c}{$\mu_\delta$} &
    \multicolumn{1}{c}{Sample Volume}\\
     &
    \multicolumn{1}{c}{[deg]} &
    \multicolumn{1}{c}{[deg]} &
    \multicolumn{1}{c}{[pc]} &
    \multicolumn{1}{c}{[Myr]} &
    \multicolumn{1}{c}{[Myr]} &
   &
    \multicolumn{1}{c}{[mag]} & 
    &
    \multicolumn{1}{c}{[mas/yr]} &
    \multicolumn{1}{c}{[mas/yr]} &
    \multicolumn{1}{c}{[$10^6 \textrm{pc}^3$]} \\
    
     \hline 
    
     Pleiades &  56.750 & $ 24.117$ &   136 &  130 & 128 & $-0.01$ & 0.045 & 1059 & $ 19.997$ & $-45.548$ & 7.4 \\
 alpha Persei &  51.675 & $ 48.800$ &   175 &  100 &  81 & $ 0.14$ & 0.090 &  598 & $ 22.929$ & $-25.556$ & 6.4 \\
    NGC 2451A & 115.800 & $-38.400$ &   193 &   60 &  50 & $-0.08$ & 0.000 &  311 & $-21.063$ & $ 15.378$ & 6.4 \\
      IC 2391 & 130.133 & $-53.033$ &   152 &   50 &  43 & $-0.01$ & 0.030 &  254 & $-24.927$ & $ 23.256$ & 6.7 \\
      IC 2602 & 160.742 & $-64.400$ &   152 &   50 &  42 & $-0.02$ & 0.031 &  391 & $-17.783$ & $ 10.655$ & 6.7 \\
 \end{tabular}
    \label{tab:sample}
\end{table*}

\begin{figure}
    \centering
    \includegraphics[width=\columnwidth,clip,trim=0 0.4in 0 0.4in]{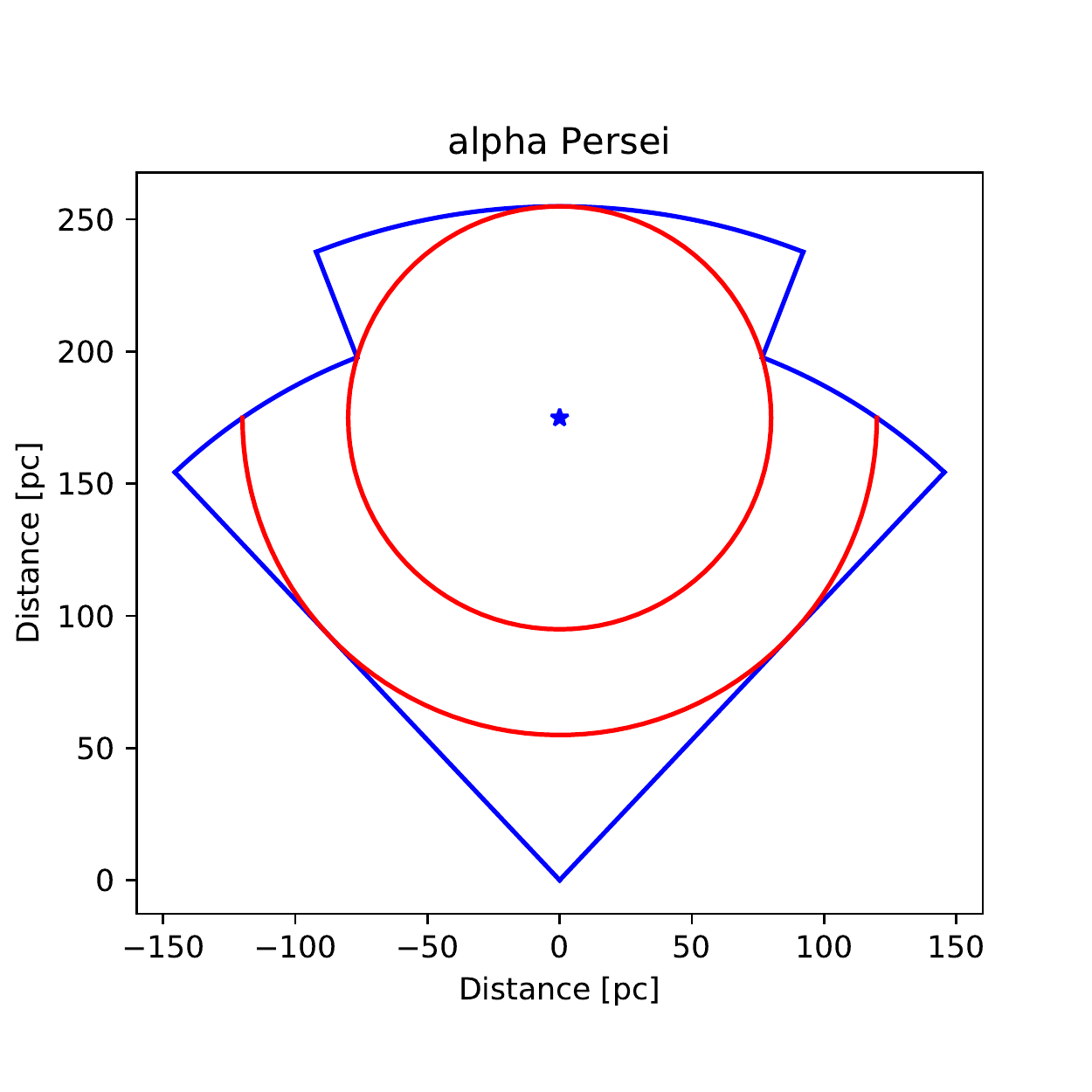}
    \caption{An illustration of the sample volume for the open cluster Alpha Persei.  The full volume within 80~pc of the centre of the cluster and the full nearside hemisphere within 120~pc are sampled.}
    \label{fig:sample_volume}
\end{figure}

\begin{figure*}
    \centering
    \includegraphics[width=0.32\textwidth,trim=0 0.2in 0 0]{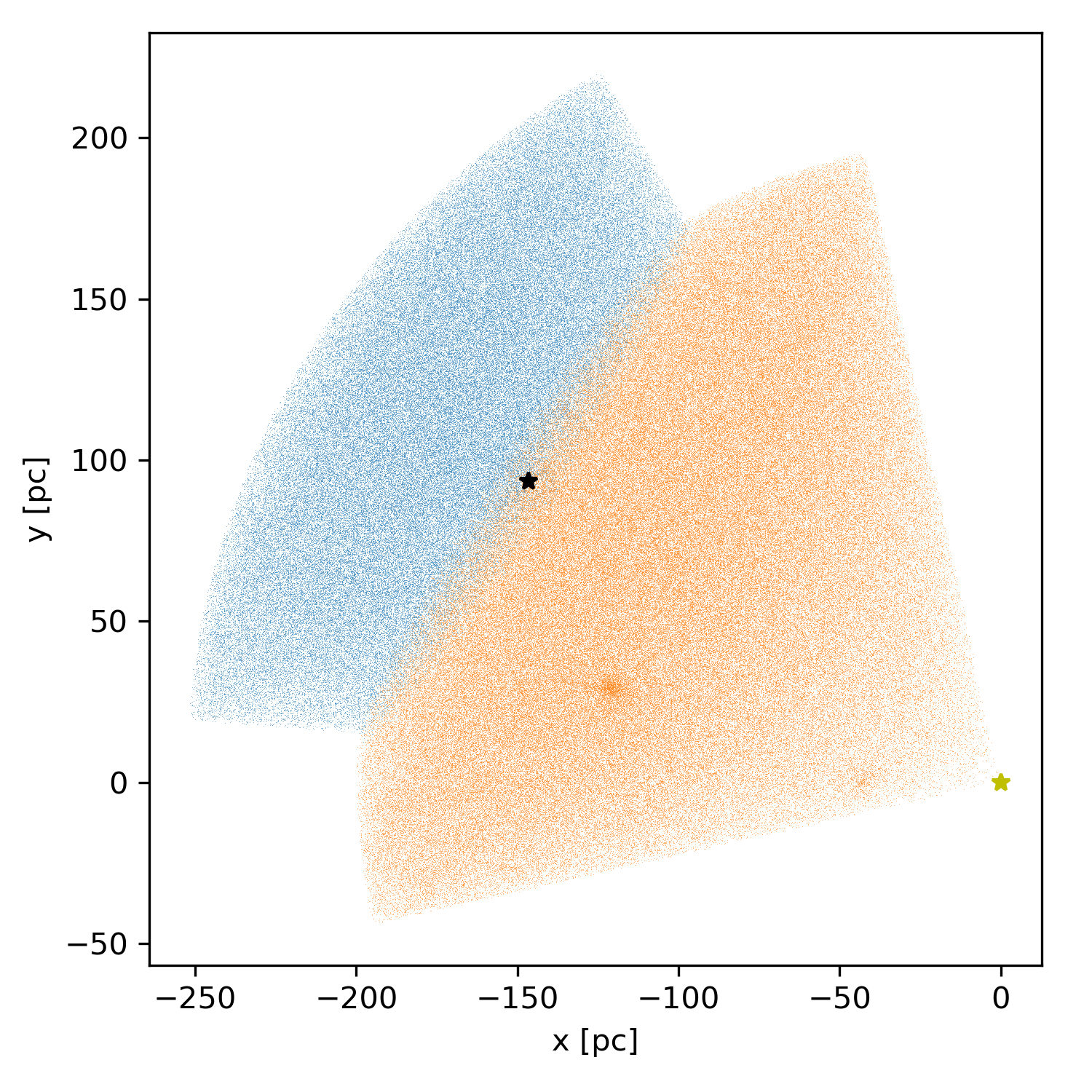}
    \includegraphics[width=0.32\textwidth,trim=0 0.2in 0 0]{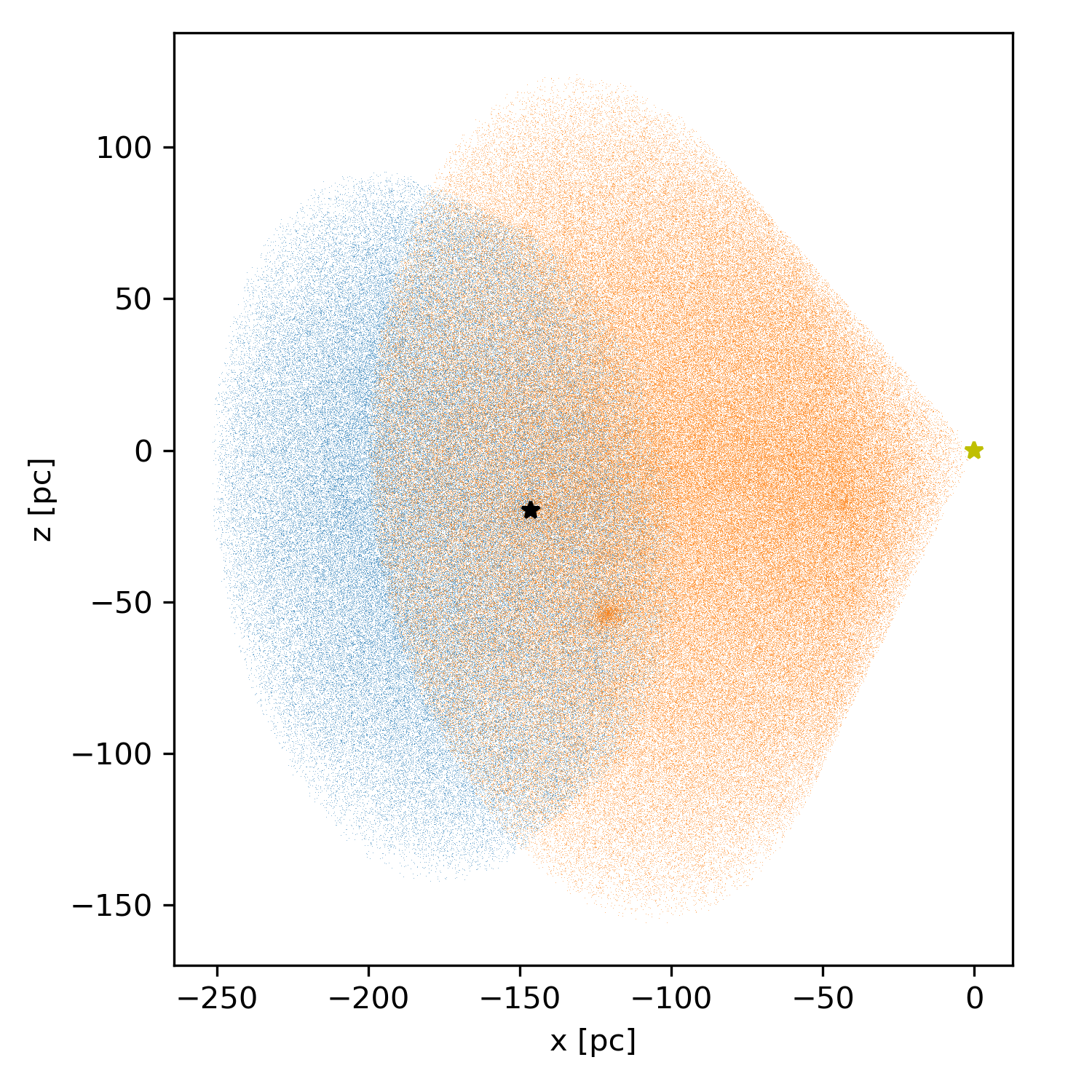}
    \includegraphics[width=0.32\textwidth]{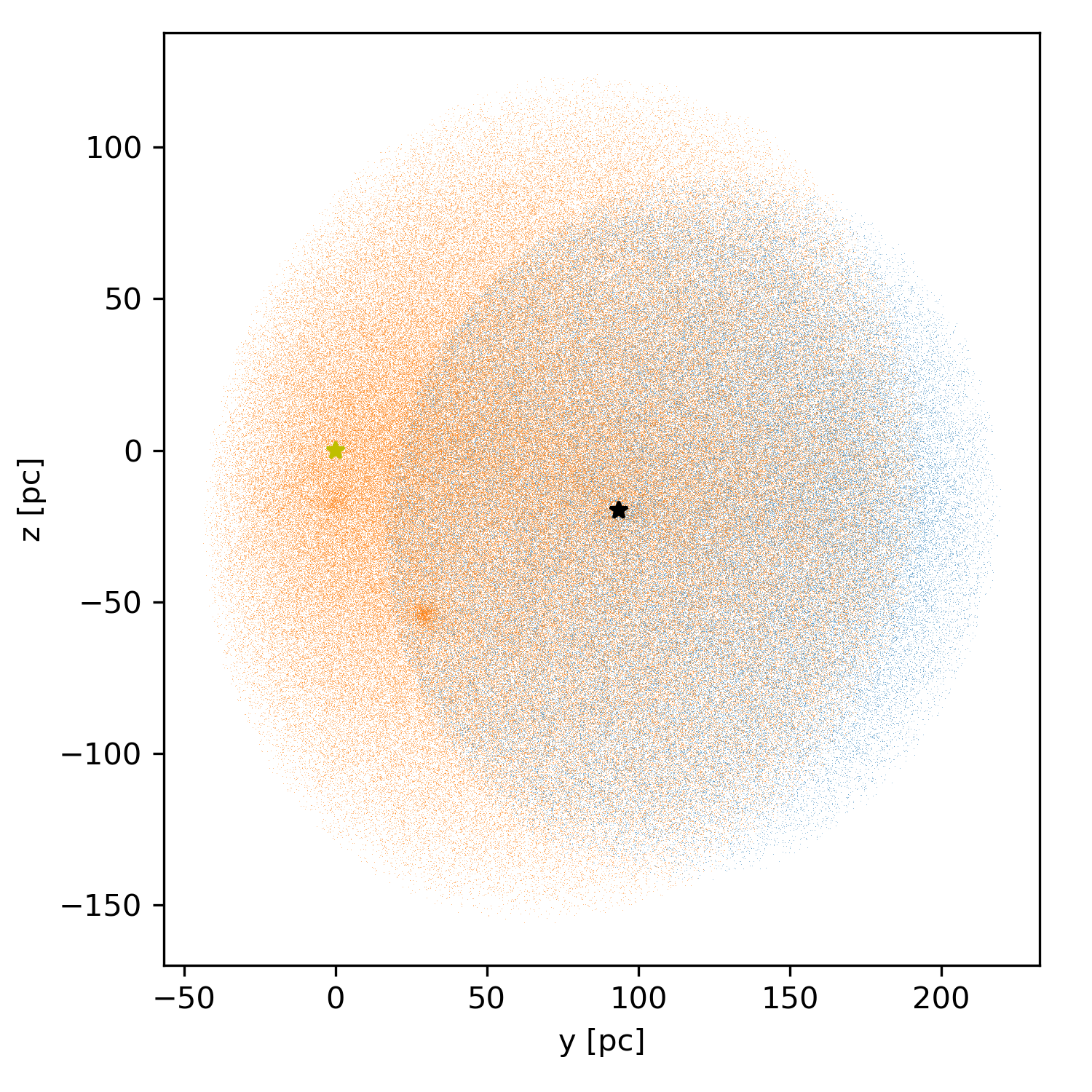}
    
    \caption{The Gaia EDR3 sample.  The blue objects lie beyond the plane containing the alpha Persei cluster and perpendicular to the line of sight from the Sun.}
    \label{fig:positions}
\end{figure*}

We will illustrate how we construct the samples with the rich young cluster, alpha Persei.  The particulars of the Gaia queries for each cluster are given in the appendix. We construct our Gaia EDR3 sample for the alpha Persei cluster by finding all objects within 250~pc of the Sun that lie within 28~degrees of the alpha Persei cluster on the sky and a second region within 200~pc of the Sun and within 45~degrees of the cluster on the sky (Fig.~\ref{fig:sample_volume}).  This results in a seven-million-cubic parsec volume as shown in Fig.~\ref{fig:positions}.  To consider the sampled volumes in greater detail we consider a region beyond the cluster (blue) and closer (orange).  From the point of view of the alpha Persei cluster itself, which is about 175~pc away from the Sun, the sample contains the entire volume within 60~pc of the cluster as well as an entire hemisphere of radius 90~pc on the nearside (by design).

\section{Methods}
\label{sec:methods}

\begin{figure}
    \centering
    \includegraphics[width=\columnwidth]{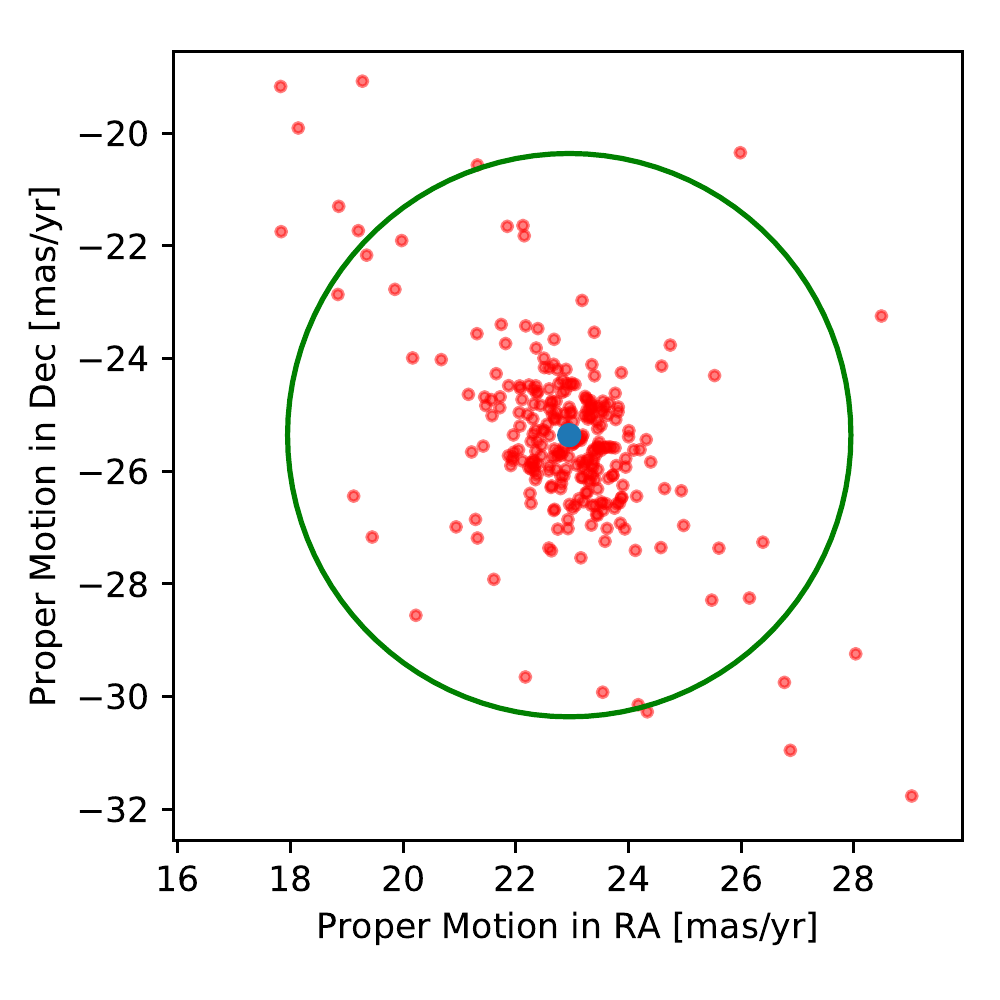}
    \caption{Proper motion of stars within one degree of the centre of the alpha Persei cluster.  The median proper motion and the five milliarcsecond per year selection region are also depicted.}
    \label{fig:pmselection}
\end{figure}

\begin{figure}
    \centering
    \includegraphics[width=\columnwidth]{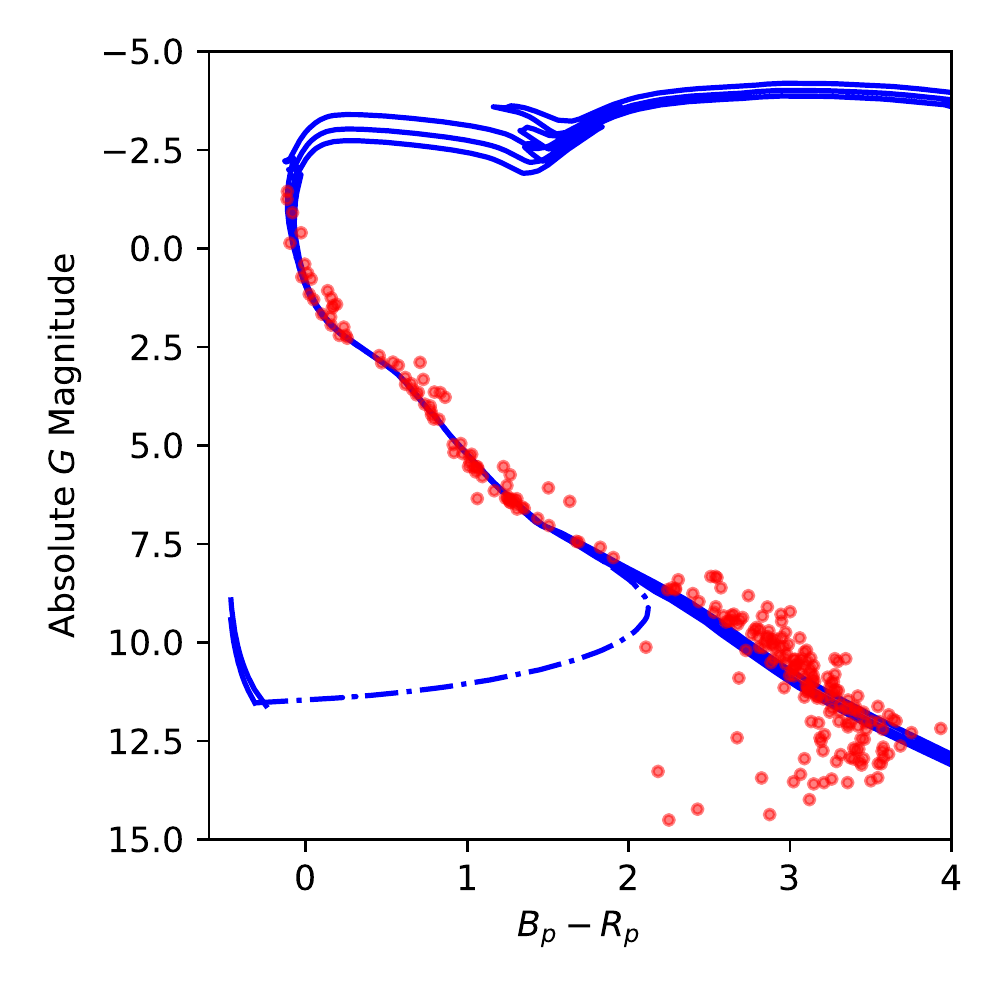}
    \caption{Gaia EDR3 colour-magnitude diagram of the proper-motion selected sample.  The blue curves are Padova isochrones ($Z=0.021$) with ages of 80, 100 and 120 and Montreal white-dwarf models of 1.0 and 1.1 solar masses up to a cooling age of 70~Myr.  An extinction of $E(B_p-R_p)=0.12$ and $A_G=0.24$ has been applied to the models.  To convert the apparent $G$ magnitudes to absolute, we use the median distance of the sample of 174~pc obtained from the parallaxes. A binary composed of the oldest white dwarf expected in the cluster and a main-sequence star is shown by the dot-dashed curve.}
    \label{fig:cmd=pmselection}
\end{figure}

We follow the same techniques as outlined in Paper I, and we will recapitulate them here for the particular case of alpha Persei.  The procedure for the other young clusters are slightly different, and we will outline these differences and the reasons behind them in \S~\ref{sec:young-clusters}. To find the centre of mass motion and position of the alpha Persei cluster, we examine the proper motions of all stars within one degree on the sky of the centre of the cluster as shown in Fig.~\ref{fig:pmselection}. We iteratively calculate the median proper motion of the cluster by starting with an initial guess of $(19, -45)~\textrm{mas/yr}$ and considering only the proper motions within five milliarcseconds per year of the median yielding the median proper motion and sample depicted in the figure.  The colour-magnitude diagram of this sample is depicted in Fig.~\ref{fig:cmd=pmselection}.  We see that the cluster is well characterized by Padova isochrones \citep{2012MNRAS.427..127B,2014MNRAS.445.4287T,2014MNRAS.444.2525C,2015MNRAS.452.1068C,2017ApJ...835...77M,2019MNRAS.485.5666P,2020MNRAS.498.3283P} from 80 to 120~Myr.  We have also plotted Montreal white-dwarf cooling models \citep{1995PASP..107.1047B,2006ApJ...651L.137K,2006AJ....132.1221H,2011ApJ...730..128T,2011ApJ...737...28B,2018ApJ...863..184B,2020ApJ...901...93B} up to a cooling age of 110~Myr on the colour-magnitude diagram. 

Using those stars within the sample that have radial velocities measured with Gaia DR2 we calculate the mean velocity of the stars within the alpha Persei cluster relative to the Sun as
\begin{equation}
 {\bf v}_\textrm{cluster} = (-13.9\pm 0.8, -24.2 \pm 0.4,-6.83 \pm 0.2 )~\textrm{km/s}
 \label{eq:1}
 \end{equation}
 in Galactic coordinates.  Using all of the stars within the sample, we calculate the mean displacement of the cluster relative to the Sun as 
 \begin{equation}
 {\bf r}_\textrm{cluster} = (-146.5\pm 0.7, 93.5 \pm 0.4,   -19.9 \pm 0.1)~\textrm{pc},
 \label{eq:2}
 \end{equation}
 again in Galactic coordinates.  Both of these values agree within uncertainties with those derived by \citet{2019A&A...628A..66L} from Gaia DR2 data.
 
 Having determined the centre of the cluster in phase space, we then used a more relaxed definition of the cluster sample to be all of the stars within 10~pc of the cluster centre whose proper motions lie within five milliarcseconds per year of the cluster median.  This sample spans about eight degrees on the sky and is depicted in Fig.~\ref{fig:cmd}.

To look for potential escapees from the alpha Persei cluster we calculate the relative velocity of all of the stars in the sample within the plane of the sky with respect to the cluster as 
\begin{equation}
\Delta {\bf v}_\textrm{2D} = {\bf v}_\textrm{2D} - {\bf v}_\textrm{cluster} + \frac{ {\bf v}_\textrm{cluster} \cdot {\bf r}}{{\bf r} \cdot {\bf r}}{\bf r},
\label{eq:3}
\end{equation}
where ${\bf v}_\textrm{2D}$ is the velocity of the star assuming zero radial velocity (see Appendix~A for the calculation).  This yields a sample of common-proper-motion objects.  If we assume that stars escape from the cluster with a velocity of only a few kilometres per second and have not been accelerated subsequently, a small value of $\Delta {\bf v}_\textrm{2D}$ is a necessary but not sufficient condition to be a candidate escapee.  

In addition to having a small proper motion relative to the cluster an object must also have a relative proper motion away from the alpha Persei cluster that is sufficiently large that it could have reached its location in the sky within the lifetime of the cluster.  To find this sample, we determine the distance of the star from the cluster as a function of time assuming no acceleration and an arbitrary radial displacement ($\delta r$)
\begin{equation}
d(t)^2 = \left [ {\bf r}-{\bf r}_\textrm{cluster} 
+  t \left ({\bf v}_\textrm{2D}-{\bf v}_\textrm{cluster} \right ) + \hat{\bf{r}} \delta r  \right ]^2,
\label{eq:4}
\end{equation}
and we look for the time when the star and the cluster were or will be closest together,
\begin{equation}
t_\textrm{min} = \frac{ \Delta {\bf r}\cdot  \Delta {\bf v} - \left (  \Delta {\bf r} \cdot \hat{\bf{r}}  \right ) \left (  \Delta {\bf v} \cdot \hat{\bf{r}}  \right ) }{\left (  \Delta {\bf v} \cdot \hat{\bf{r}}  \right )^2-\left (  \Delta {\bf v}\right )^2}
\label{eq:5}
\end{equation}
where
\begin{equation}
 \Delta {\bf r} = {\bf r}-{\bf r}_\textrm{cluster} ~\textrm{and}~
 \Delta {\bf v} = {\bf v}_\textrm{2D}-{\bf v}_\textrm{cluster}.
 \label{eq:6}
 \end{equation}
This also yields an estimate of the radial displacement and velocity of the star
\begin{equation}
\delta r = v_r t_\textrm{min} =  -\hat{\bf{r}} \cdot \left ( \Delta {\bf r} +t_\textrm{min} \Delta {\bf v}   \right ) 
\label{eq:7}
\end{equation}
 so
 \begin{equation}
 \hat {\bf v}_\textrm{3D} = {\bf v}_\textrm{2D} + v_r \hat{\bf{r}}
 \label{eq:8}
 \end{equation}
 and
 \begin{equation}
 \Delta \hat {\bf v}_\textrm{3D} = \hat {\bf v}_\textrm{3D} - {\bf v}_\textrm{cluster}
 \label{eq:9}
 \end{equation}
 where the caret denotes that this is the reconstructed velocity, rather than the measured velocity which is also available for a subset of the sample (see Fig.~\ref{fig:rvcomp}).  We do not account for the measurement errors in the parallax, position and proper motion in this analysis.   The typical relative uncertainties on these quantities are less than a few percent for the stars in the sample. For the cluster members themselves, the median distance error is 1.7~pc and the median proper motion error in one-dimension is 0.06~milliarcseconds per year.  To place the proper motion error in context, this is 41~m/s at the distance of the alpha Persei cluster, and over the course of 80~Myr, this would result in a position error of 4~pc.  These two distance errors are well within the threshold value of $d_\textrm{min}$ that we will use to identify potential escapees (15~pc). 
 
\begin{figure}
    \centering
    \includegraphics[width=\columnwidth]{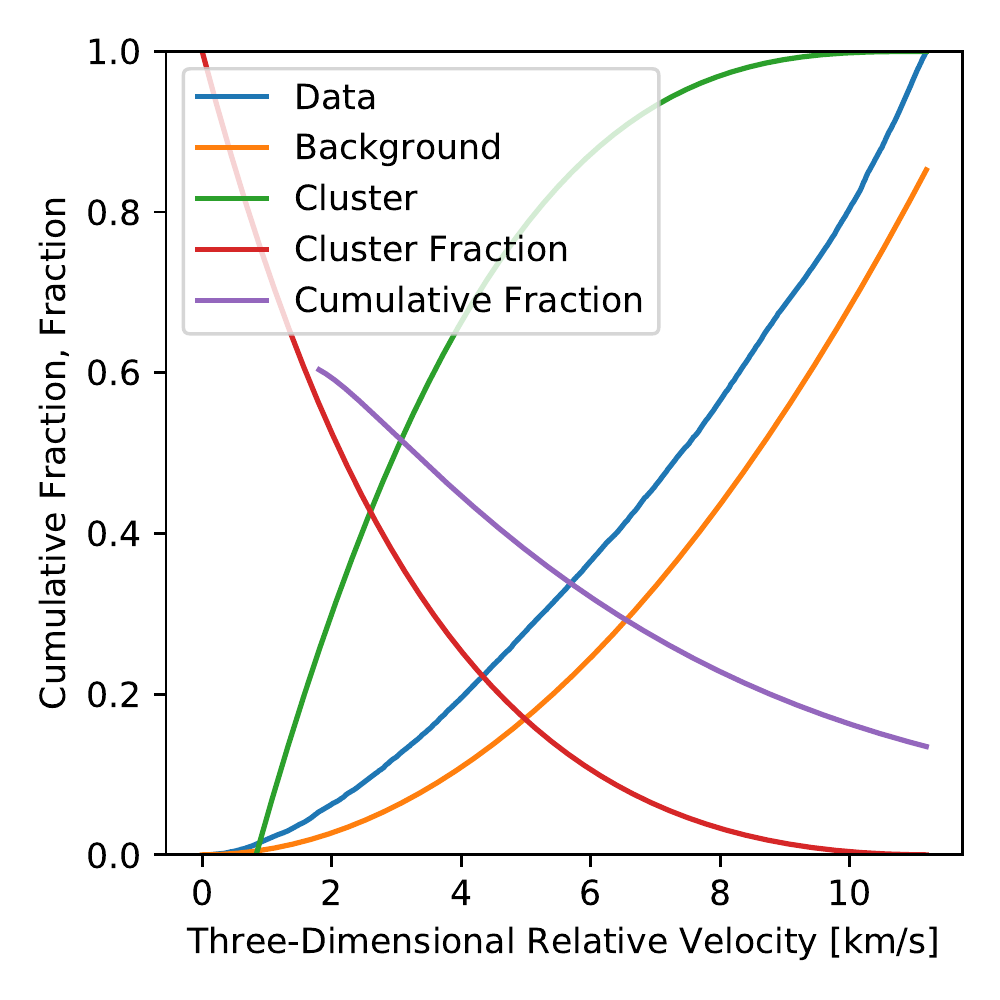}

    \caption{The cumulative distribution of reconstructed velocities for the stars that have come within 15~pc of the alpha Persei cluster (blue).  There is an excess of stars at small velocities above the background (orange) which we model as uniform in two dimensions. The cumulative distribution of the excess over the background, the cluster, is shown in green.  The fraction of stars at a given velocity that are cluster members is given in red, and the cumulative fraction within a given velocity is depicted in purple. }
    \label{fig:frac3d}
\end{figure}

\begin{figure}
    \centering
    \includegraphics[width=\columnwidth]{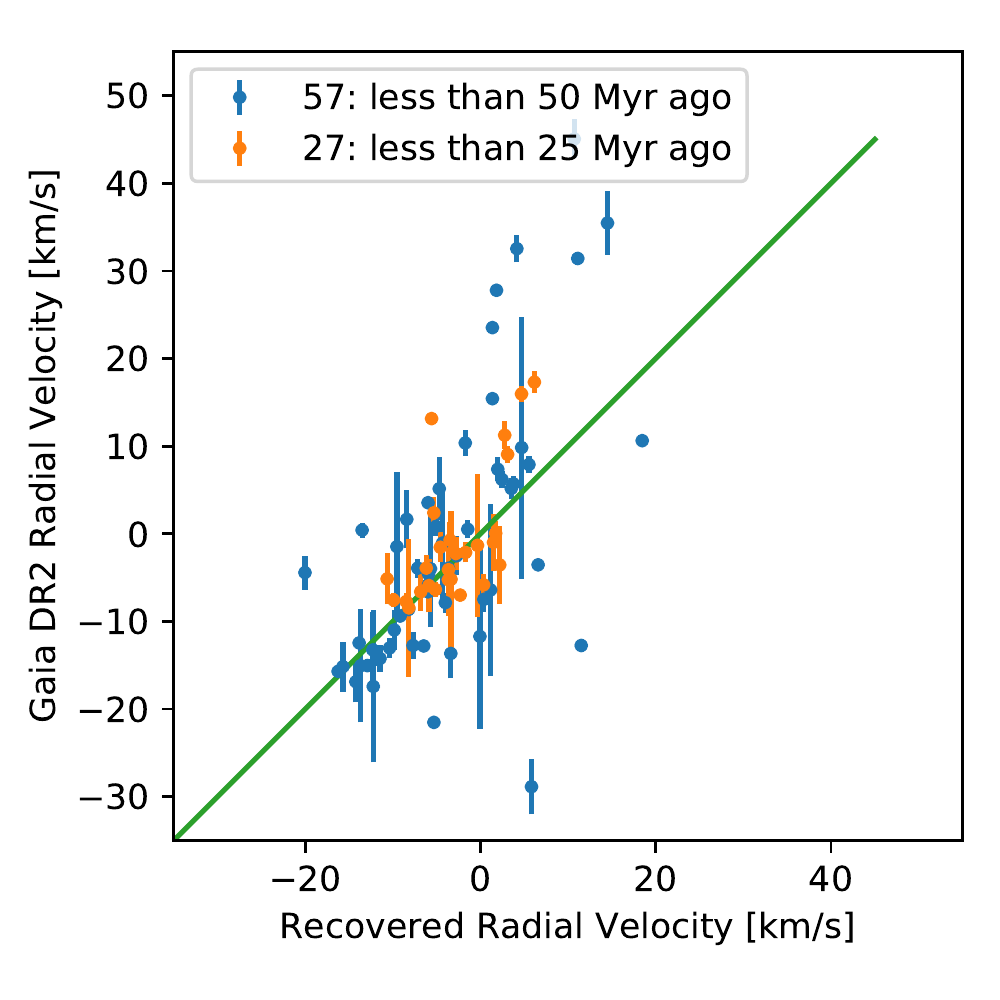}
    \caption{A comparison of Gaia DR2 radial velocities to the reconstructed radial velocities for escapee candidates.  The single escapee candidate with a Gaia DR2 radial velocity with an escape time greater than 50~Myr of 126~Myr has been omitted. }
    \label{fig:rvcomp}
\end{figure}

 To determine the threshold for the relative velocity to be deemed a candidate escapee we look at the cumulative distribution of the magnitude of the three-dimensional reconstructed relative velocity $\Delta \hat {\bf v}_\textrm{3D}$, as shown by the blue curve in Fig.~\ref{fig:frac3d}.  We have restricted the sample to relative velocities less than 11~km/s and those stars that were less than 15~pc from the centre of the cluster in the past 100~Myr which we have taken as an estimate of the age of the cluster \citep{2018A&A...616A..10G}.   With these constraints in mind, we can see that the magnitude of the reconstructed radial velocity $v_r$ is strongly correlated with the magnitude of $\Delta {\bf v}$, so we expect the relative velocity of the background to be uniformly distributed in two dimensions as shown by the orange curve which is parallel to the blue curve for large relative velocities.  With the background normalisation at large relative velocity, we can determine the cumulative fraction of cluster stars which we take to be the excess over the background (shown in green).  This yields the probability that a star with a given relative velocity is a member of the cluster distribution (red) and fraction of cluster stars in a sample defined by a given maximum relative velocity (purple).

In particular, we find that the distribution of the relative velocity of the escapee candidates is strongly peaked toward small velocities as expected from the relaxation of the cluster velocity distribution toward a Maxwellian distribution through two-body interactions \citep{2008gady.book.....B}.  The escaped stars typically have a velocity of $\sqrt{2}$ times the velocity dispersion of the cluster, 1.6~km/s in the case of the alpha Persei cluster; consequently we expect escapees to travel up to a few km/s relative to the cluster.  At a relative velocity of 3.06~km/s, we find that both the cumulative fraction of cluster stars and the fraction of cluster stars within this relative velocity is 52.8\%.  We choose this as our threshold for a star to be an escapee candidate as at this value the true-positive rate equals the completeness rate (cumulative fraction of cluster stars) so that the total number of stars within the escapee candidate sample is expected to equal the total number of escapees. 

Fortunately, for a sub-sample of our escapee candidates Gaia DR2 has measured radial velocities so that we can compare the reconstructed radial velocity $v_r$ (Eq.~\ref{eq:6}) with the measured value.  Fig.~\ref{fig:rvcomp} compares the reconstructed radial velocities with those observed for stars that we estimate to have left the cluster within the last fifty million years.  There is only a single star that we think left the cluster earlier with a Gaia DR2 radial velocity.  For the stars that have escaped within the last twenty-five million years there is excellent correspondence between the reconstructed and observed radial velocities with only a single outlier; therefore, we argue that the true positive rate for the recent escapers is likely to be larger than inferred from Fig.~\ref{fig:frac3d}. For the stars that left earlier, there are more outliers perhaps seven out of sixteen, meaning that 56\% of the escapees identified in this sample have reconstructed and observed radial velocities in agreement, similar to the true-positive rate shown in Fig.~\ref{fig:frac3d} of 53.5\%.

\section{Results}
\label{sec:results}

\subsection{alpha Persei}

Having defined our sample of candidate escapees, we discuss the sample in further detail including the colour-magnitude diagram  and the effects of evaporation on the cluster.  We devote a separate paper to look at the escaped white dwarfs \citep{alphaperWD}.
\begin{figure*}
    \centering
    \includegraphics[width=0.32\textwidth,trim=0 0.2in 0 0]{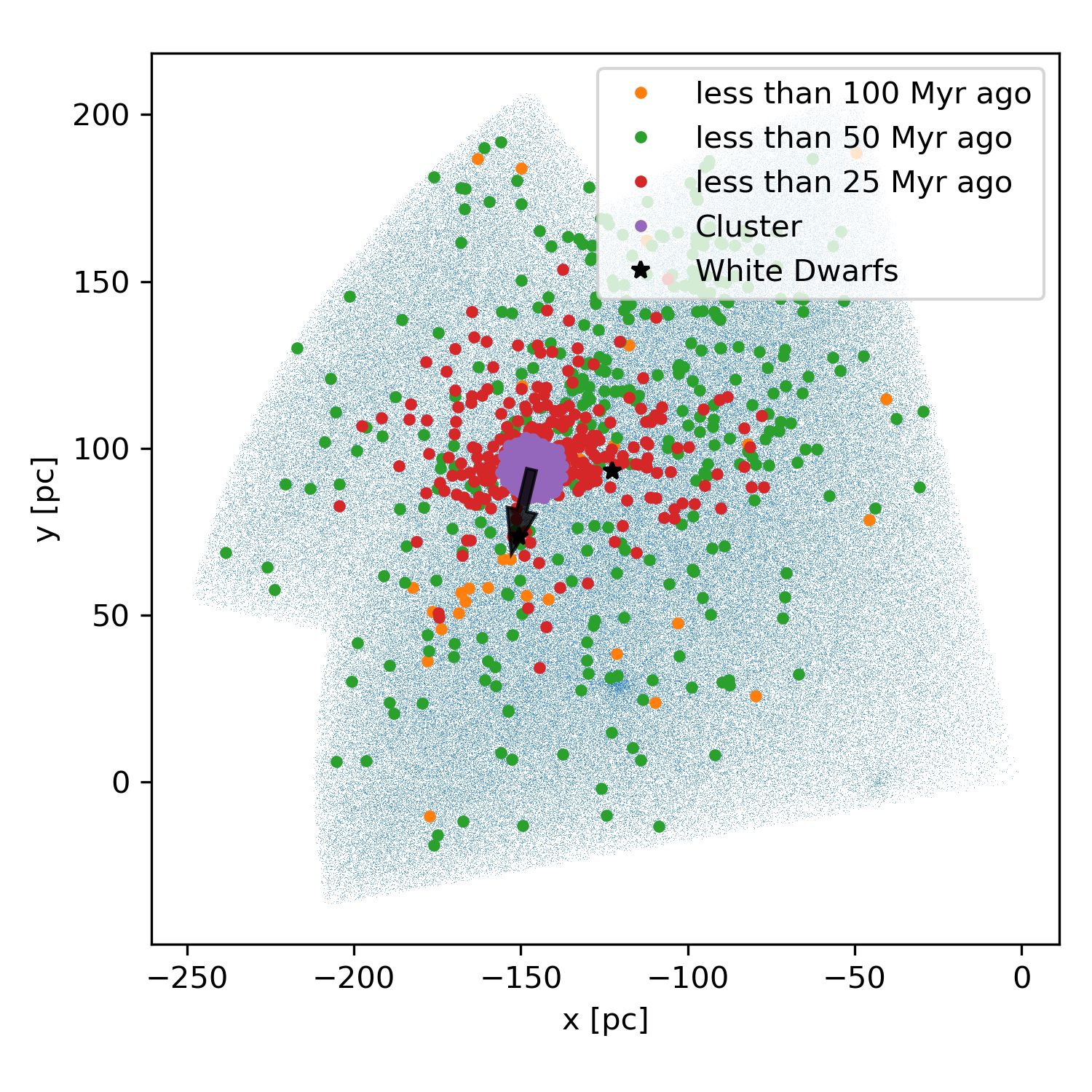}
    \includegraphics[width=0.32\textwidth,trim=0 0.2in 0 0]{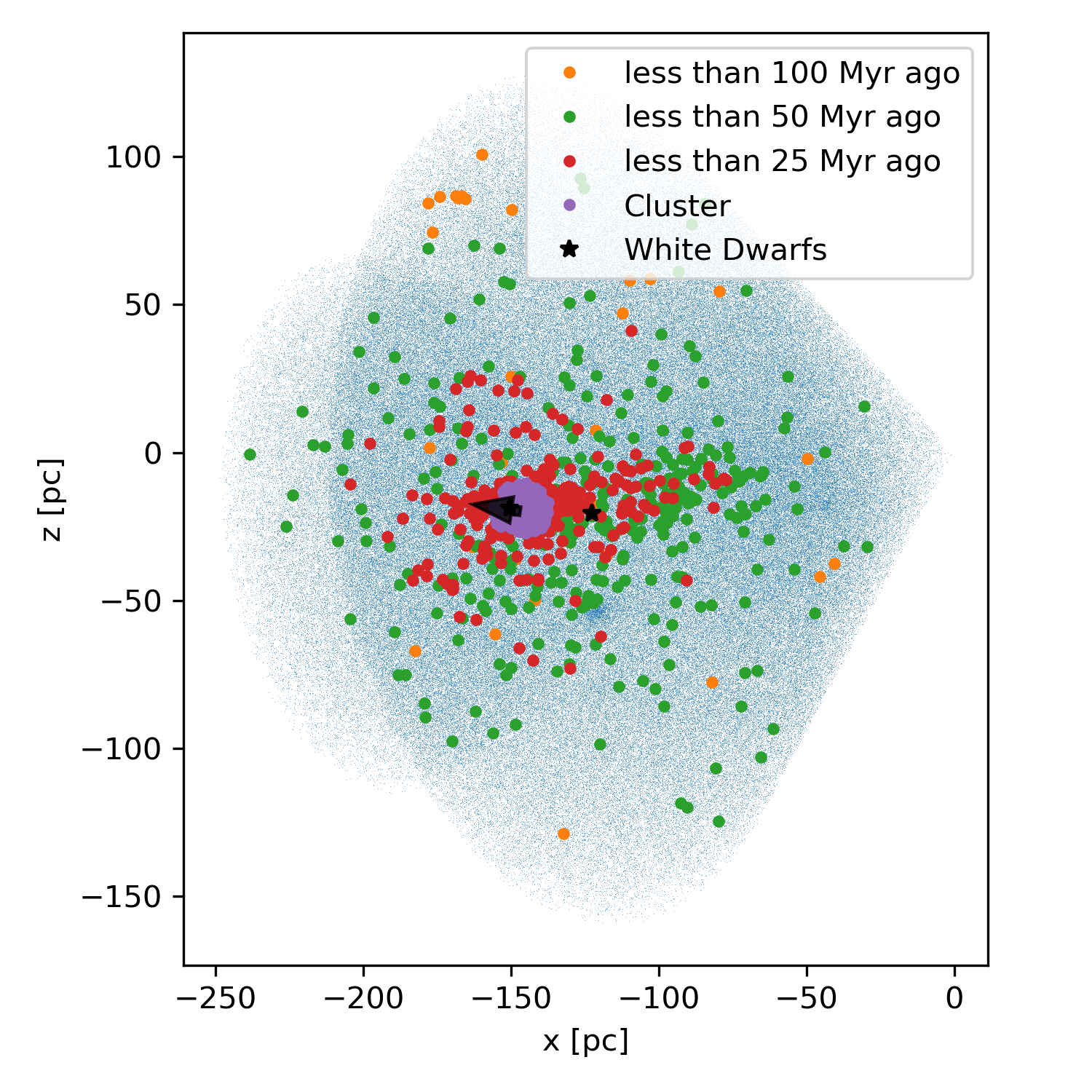}
    \includegraphics[width=0.32\textwidth]{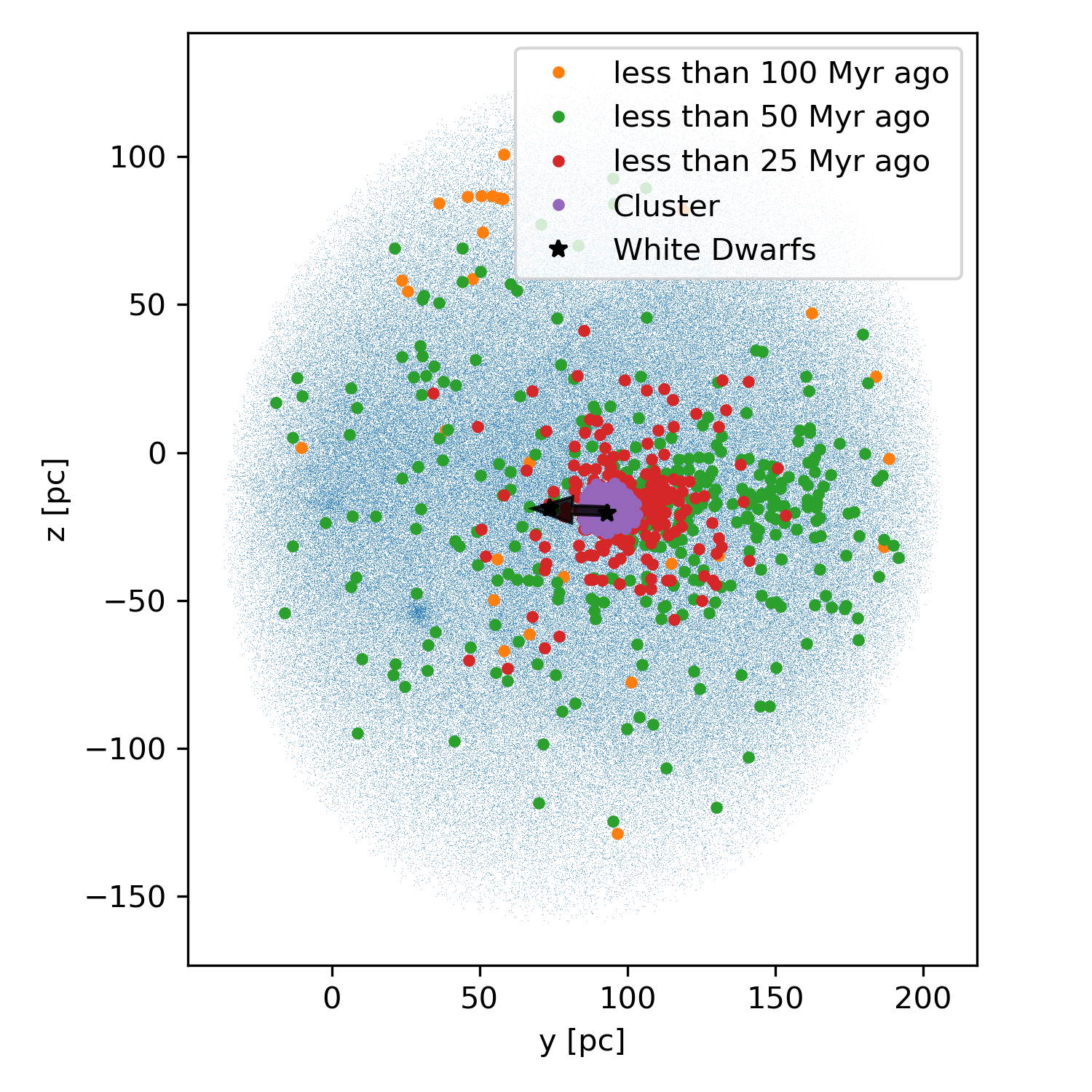}
        \caption{Where are they now?  The location of the Sun is at the origin, and the displacement of the alpha Persei cluster over one million years relative to the local standard of rest is shown by the arrow. }
    \label{fig:esc-positions}
\end{figure*}
Fig.~\ref{fig:esc-positions} shows the positions of the candidate escapees as a function of the time of escape with the cluster stars and the two white dwarfs identified as well.  The Sun is located at the origin.  Even if we account for stars that may have left the sampling volume, it is apparent that rate of evaporation from the cluster is increasing.  We have also plotted the motion of the cluster relative to the Galaxy.  To obtain the velocity of the cluster relative to the local standard of rest (LSR), we add the solar velocity relative to the LSR 
 \begin{equation}
 {\bf v}_\odot = \left (11.1^{+0.69}_{-0.75}, 12.24^{+0.47}_{-0.47}, 7.25^{+0.37}_{-0.36} \right )~\textrm{km/s}
 \end{equation}
to that of the cluster \citep{2010MNRAS.403.1829S}.  Apparently the cluster is travelling approximately parallel to the Galactic plane.

The maximum relative velocity that a star can have and still be identified as a candidate escapee is 3.07~km/s so over the course of 25~Myr an escapee can only travel 79~pc from the centre of the cluster; therefore, all of the stars that have escaped from the cluster with velocities this small must still be in the sampled volume.  To estimate the fraction of the escapees that remain in the sampling volume after longer periods, we can simply evolve the reconstructed velocities of those stars that have escaped within the last 25~Myr to simulate the older samples.  Table~\ref{tab:volume} shows the number of stars that remain within the total volume ($N_1$) and the nearside volume ($N_{1/2}$) as time progresses.  We see that the nearside volume contains a larger fraction of the escapees than the total volume.  From the geometry in Fig.~\ref{fig:positions}, this result is expected.  However, because the distribution of the velocities of the escapees is anisotropic a larger number of stars are contained with the sampled volume than would be obtained if the velocity distribution were spherically symmetric.  The final columns give the fraction of escapees that remain in the volume and, in particular, we will use the second to last column $C_1$ to correct the samples of escapees to estimate the mass function of the cluster as a function of time and the total mass of stars that have left the cluster.
\begin{table}
     \caption{Volumetric Completeness: The subscript 1 refers to the total volume, and the $1/2$ refers the nearside volume shown in orange in Fig.~\ref{fig:positions}.}
     \centering
     \begin{tabular}{c|rrrrr}
     \hline
     Escape Time & $d_\textrm{max}$ & $N_1$ & $N_{1/2}$  & $C_1$ & $C_{1/2}$ \\
     
     [Myr] & [pc] \\
     \hline
$0-25$   &  79 & 227 & 126 & 1.00 & 1.00 \\
$25-50$  & 158 & 184 & 110 & 0.81 & 0.87 \\
$50-75$  & 237 & 113 &  68 & 0.50 & 0.54 \\
$75-100$ & 315 &  68 &  41 & 0.30 & 0.33
     \end{tabular}
     \label{tab:volume}
 \end{table}
 
\begin{figure}
    \centering
    \includegraphics[width=\columnwidth]{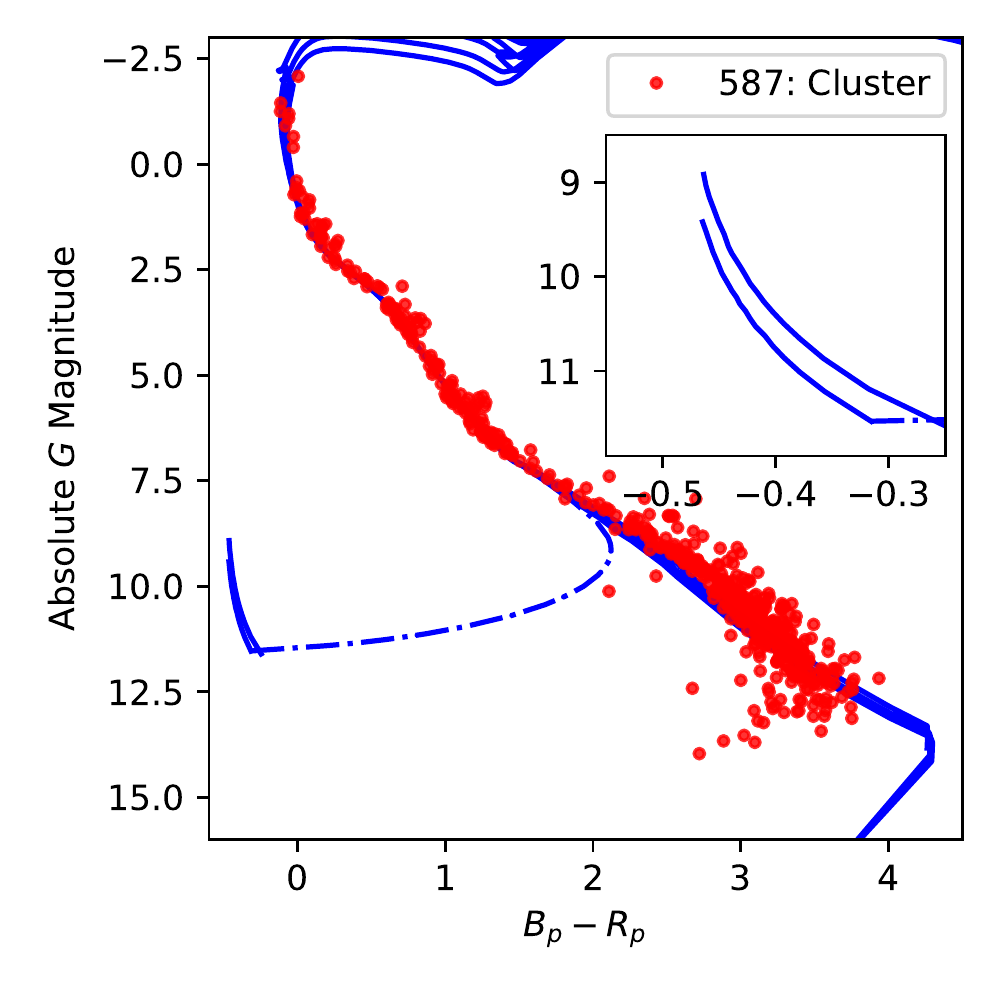}
    \caption{Colour-magnitude diagram of the alpha Persei cluster today using the volume-limited sample of those objects within 10~pc of the centre of the cluster and with proper motions lying within the range shown in Fig.~\ref{fig:pmselection}.  The blue curves are as in Fig.~\ref{fig:cmd=pmselection}.}
    \label{fig:cmd}
\end{figure}
We will first focus on the colour-magnitude diagram of the cluster stars that we define here to be those stars within 10~pc of the centre of the cluster \citep[e.g.][]{2021A&A...645A..84M} and that also have proper motions within five milliarcseconds per year of the median cluster proper motion.  This sample is depicted in Fig.~\ref{fig:cmd}.  This group of nearly 600 stars is slightly different from the sample depicted in Fig.~\ref{fig:cmd=pmselection} which had no constraint on the distance to the star but a stronger constraint on the distance on the sky from the centre of the cluster.  We identify more stars above the turnoff in this sample, indicating that a few of the turn-off stars in Fig.~\ref{fig:cmd} have measured positions that placed them beyond the one-degree cutoff.

\begin{figure*}
    \centering
    \includegraphics[width=0.32\textwidth]{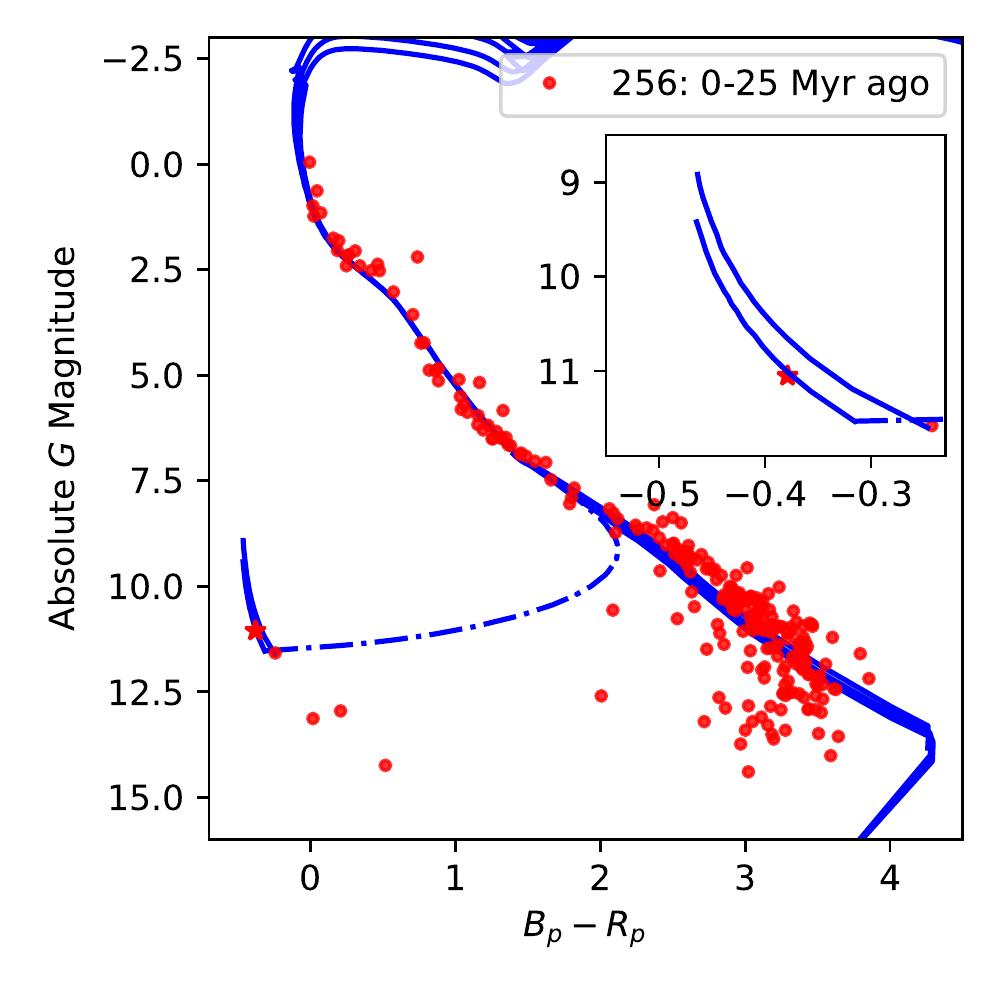}
    \includegraphics[width=0.32\textwidth]{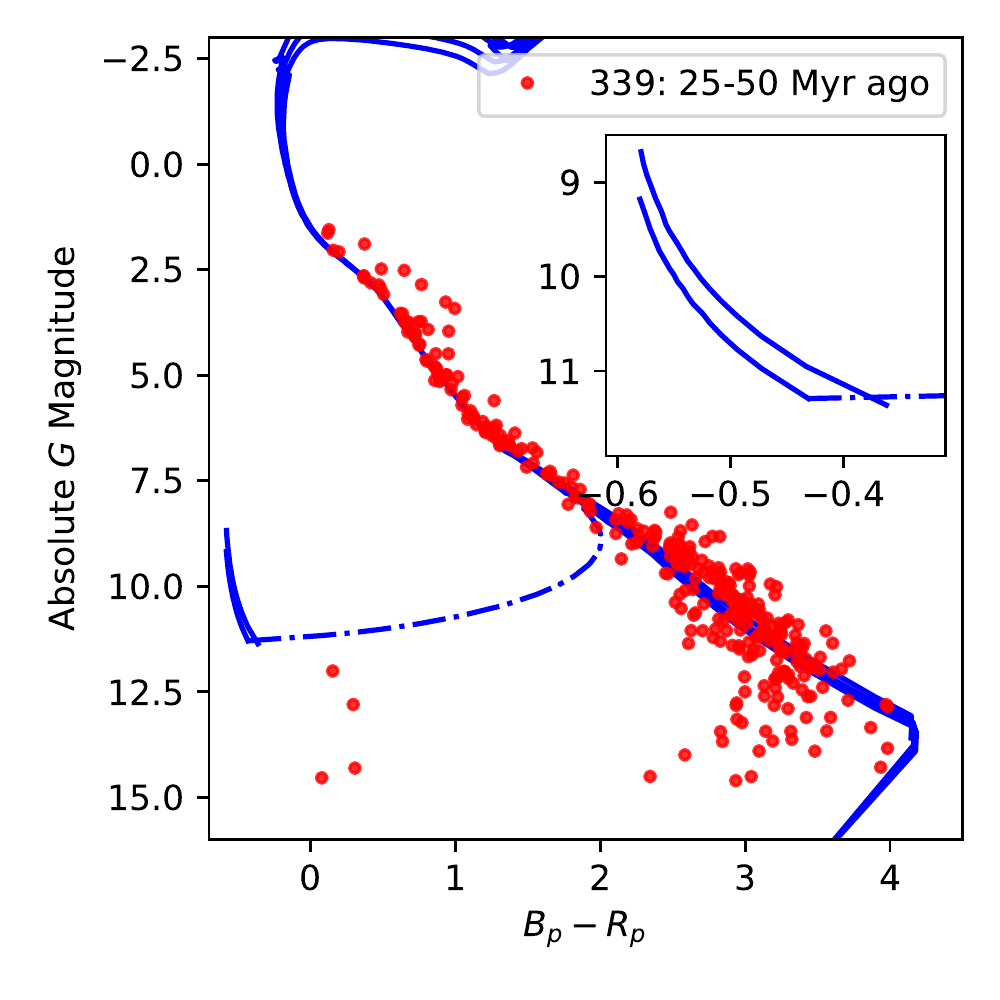}
    \includegraphics[width=0.32\textwidth]{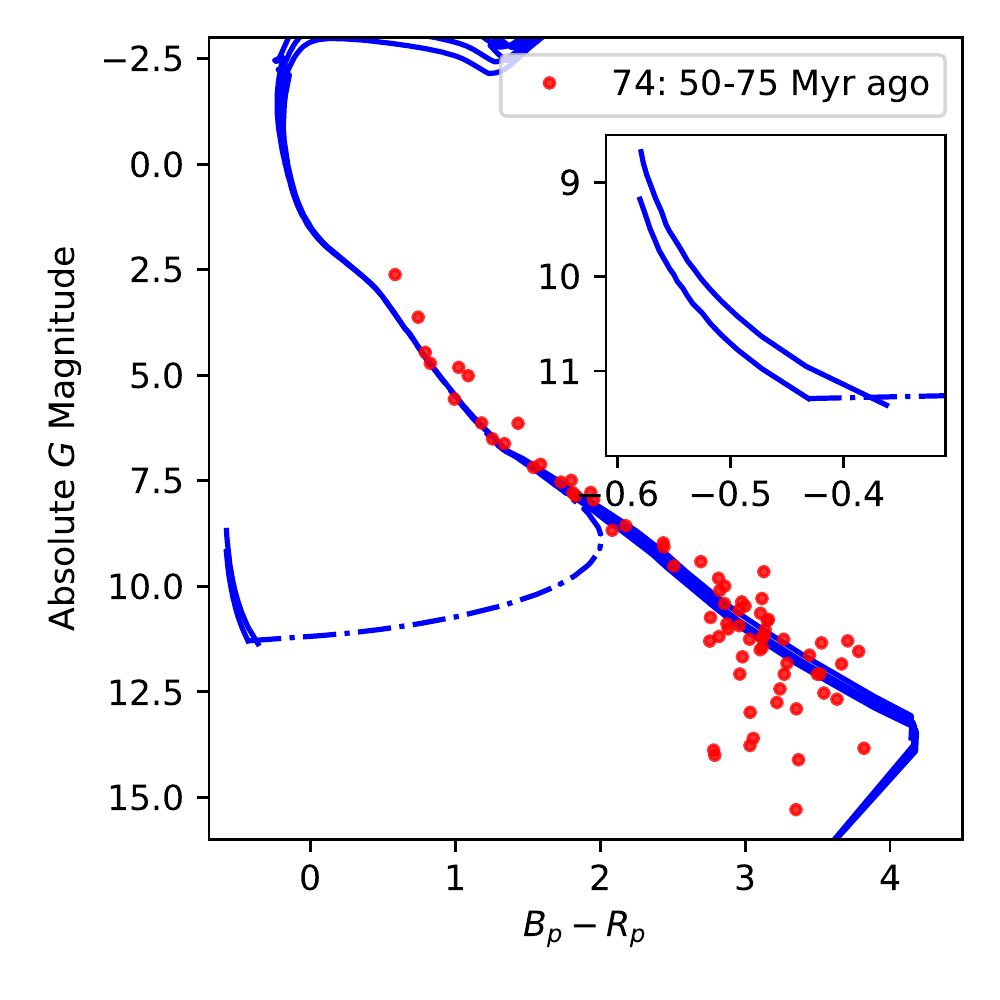}
    \includegraphics[width=0.32\textwidth]{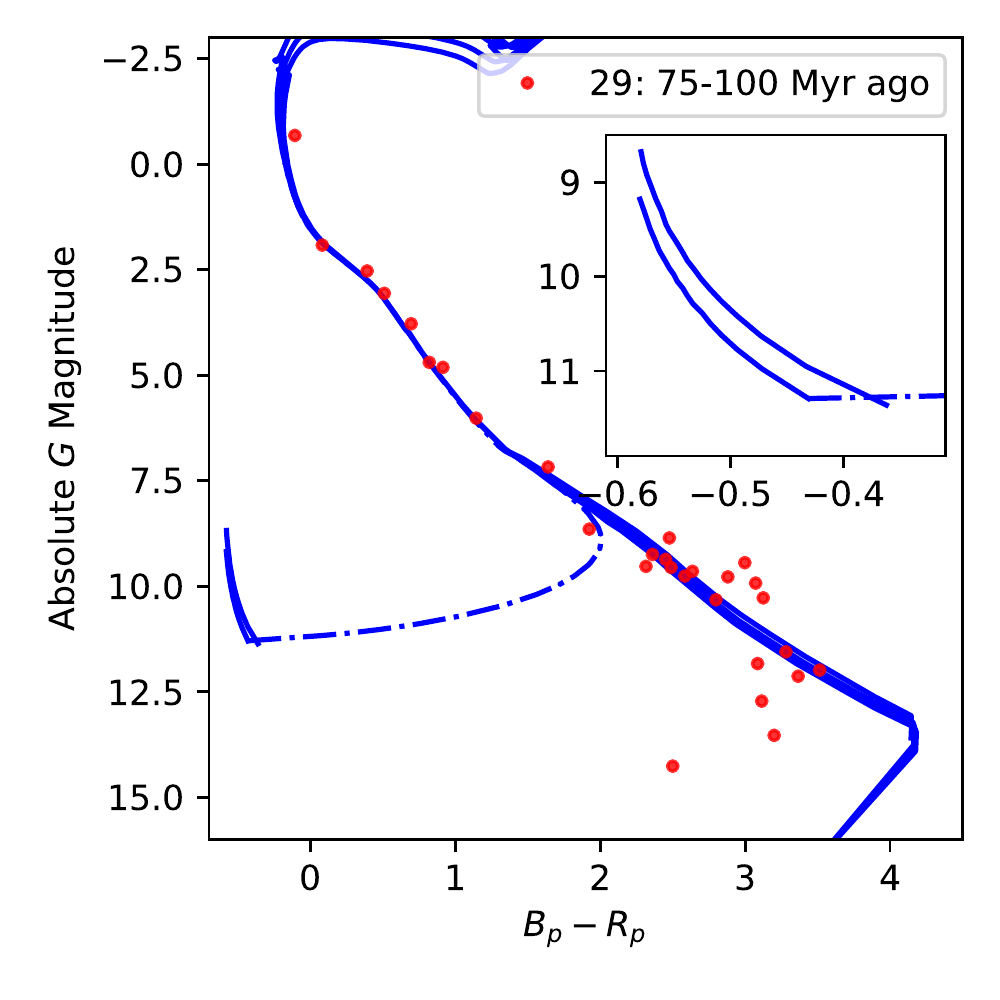}
    \includegraphics[width=0.32\textwidth]{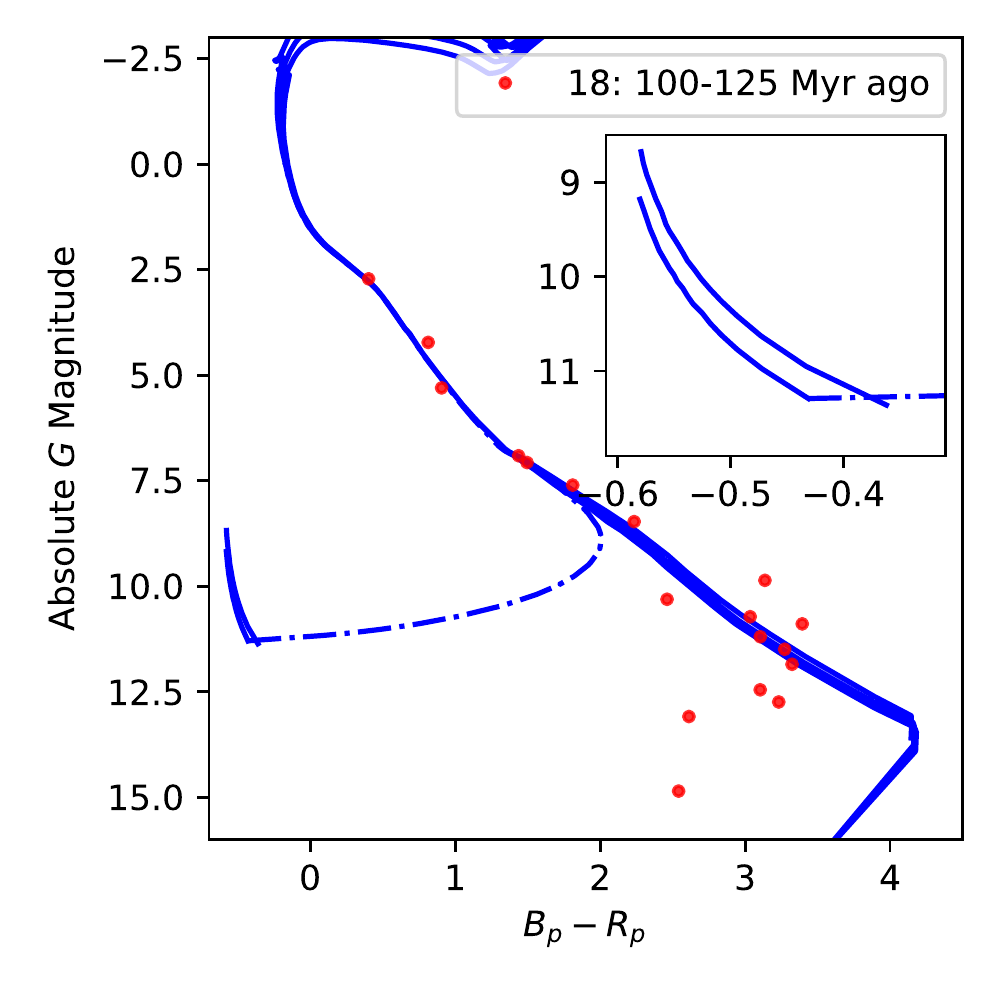}
    \includegraphics[width=0.32\textwidth]{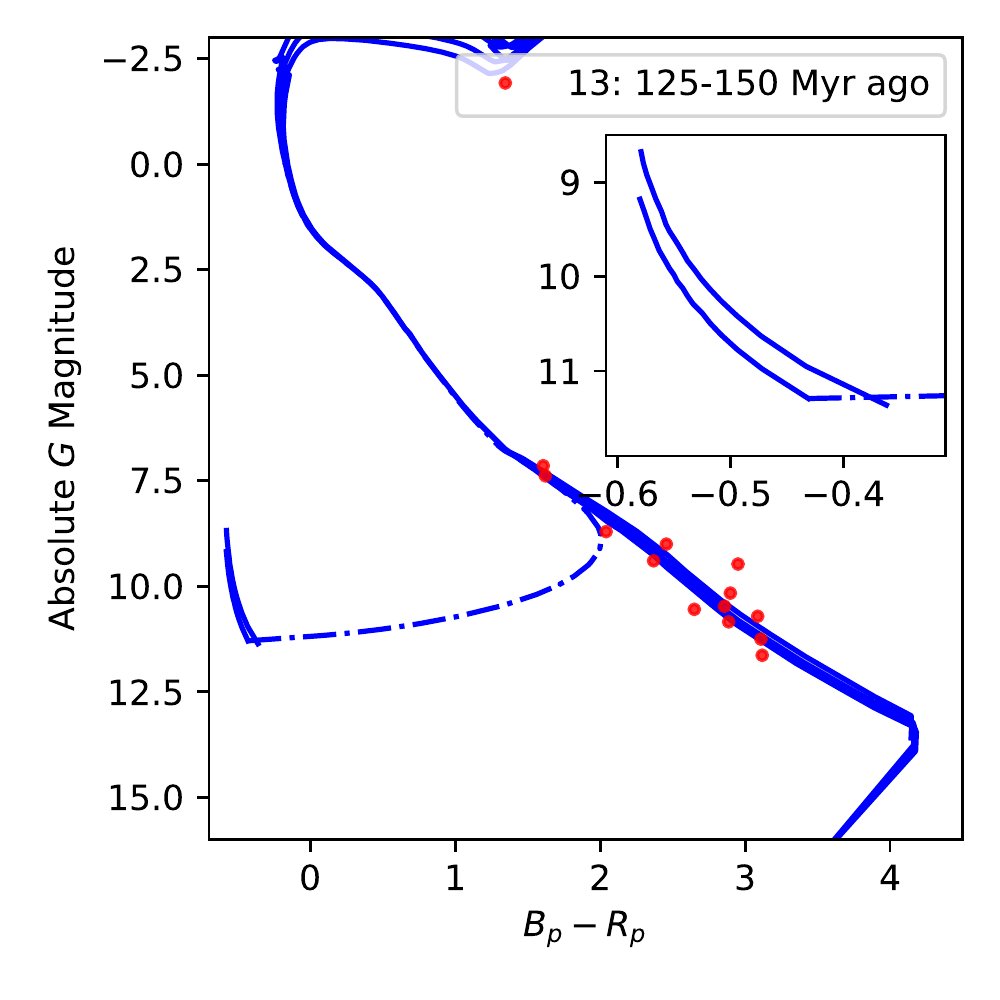}
    \caption{Colour-magnitude diagram of the alpha Persei escapees as a function of time.  The blue curves are as in Fig.~\ref{fig:cmd=pmselection} with the cluster extinction applied for the 25~Myr sample and no extinction applied for the older samples.  The individual parallaxes of the objects are used to calculate the absolute magnitudes.  The object designated by a star in the white-dwarf region left the cluster 5~Myr ago with a relative velocity of 4~km/s, larger than our cutoff velocity.  Given the rarity of young massive white dwarfs in the field we have included it in the sample.}
    \label{fig:cmd-escape}
\end{figure*}

The stellar population of the alpha Persei cluster itself is well known, so we proceed to the candidate escapees in Fig.~\ref{fig:cmd-escape}.  We first focus on the most recent escapees.  These stars lie close to the cluster in a triaxial configuration more than 10~pc from the centre of the cluster. Because the long axis of the configuration lies approximately along the line of sight, many of the stars in the 25 Myr escapee sample have already been listed as cluster members but in fact the Gaia EDR3 parallaxes place them either closer or further from the cluster.  The colour-magnitude diagram of the most recent escapees resembles that of the cluster but it is necessarily more sparse.  Among the recent escapees we can identify two massive-white-dwarf candidates that we will discuss in a follow-up paper \citep{alphaperWD} with spectroscopic observations. As we look at stars that escaped the cluster earlier, the distribution is more and more skewed toward lower-mass stars which is expected from mass segregation within the cluster  \citep{1997A&ARv...8....1M,1998MNRAS.295..691B,1998A&A...331...81P}.  In fact, we will use the properties of the early escapees to provide a kinematic estimate of the cluster age (\S~\ref{sec:k-ages}).

\begin{figure}
    \centering
    \includegraphics[width=\columnwidth]{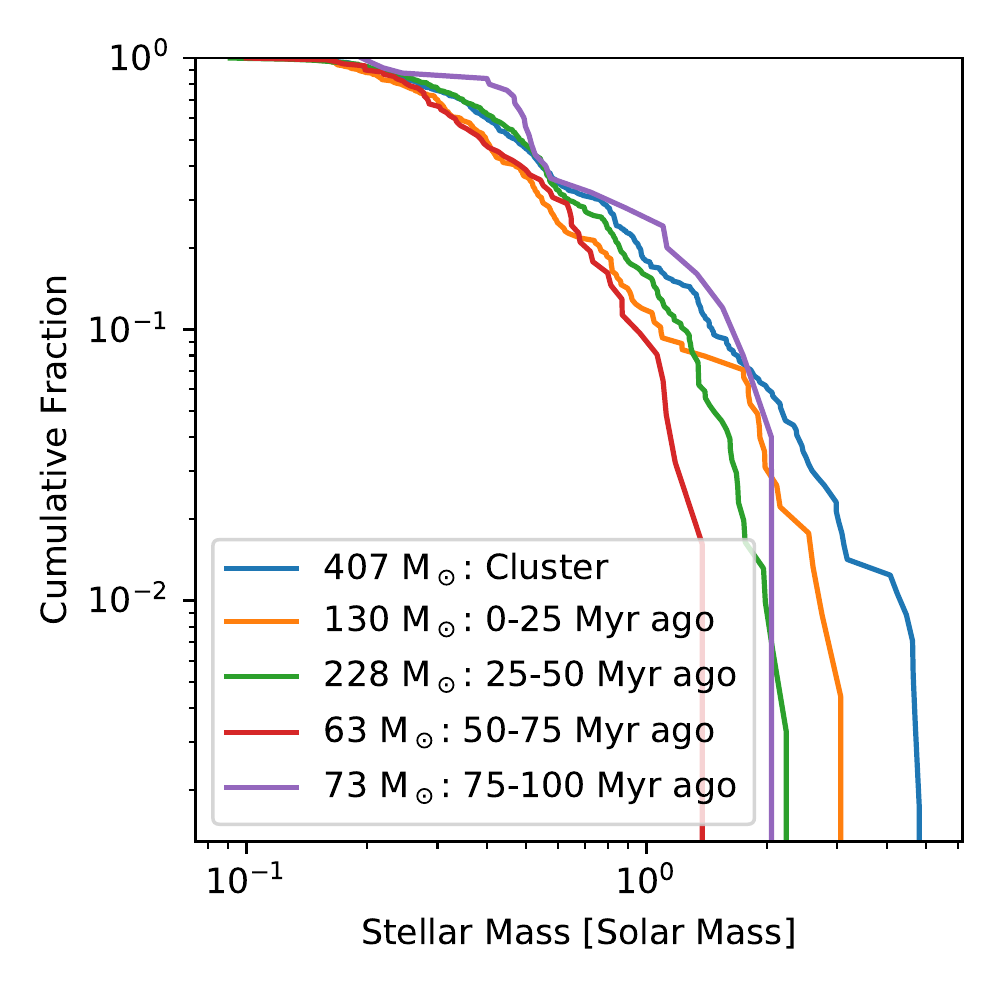}
    \caption{Cumulative Distribution of Escaped Stars as a Function of Mass}
    \label{fig:escape-mass}
\end{figure}

We can use the escaped stars depicted in the colour-magnitude diagrams in Fig.~\ref{fig:cmd-escape} to estimate the mass function of the escaped stars and how the mass function of the cluster has evolved.  To do this we focus on the stars that lie along the main sequence and use the 80 Myr Padova isochrone to infer the initial mass of each star.  The resulting cumulative mass distributions are depicted in Fig.~\ref{fig:escape-mass}. The total mass given in the legend includes a correction for the escapees that have left the sample region from Tab.~\ref{tab:volume}.  This correction is greater than a factor of two for stars that have left the cluster more than 100~Myr ago.  Because of this and the small numbers of early escapees, we only consider stars that have escaped within the last 100~Myr.  

\begin{figure}
    \centering
    \includegraphics[width=\columnwidth]{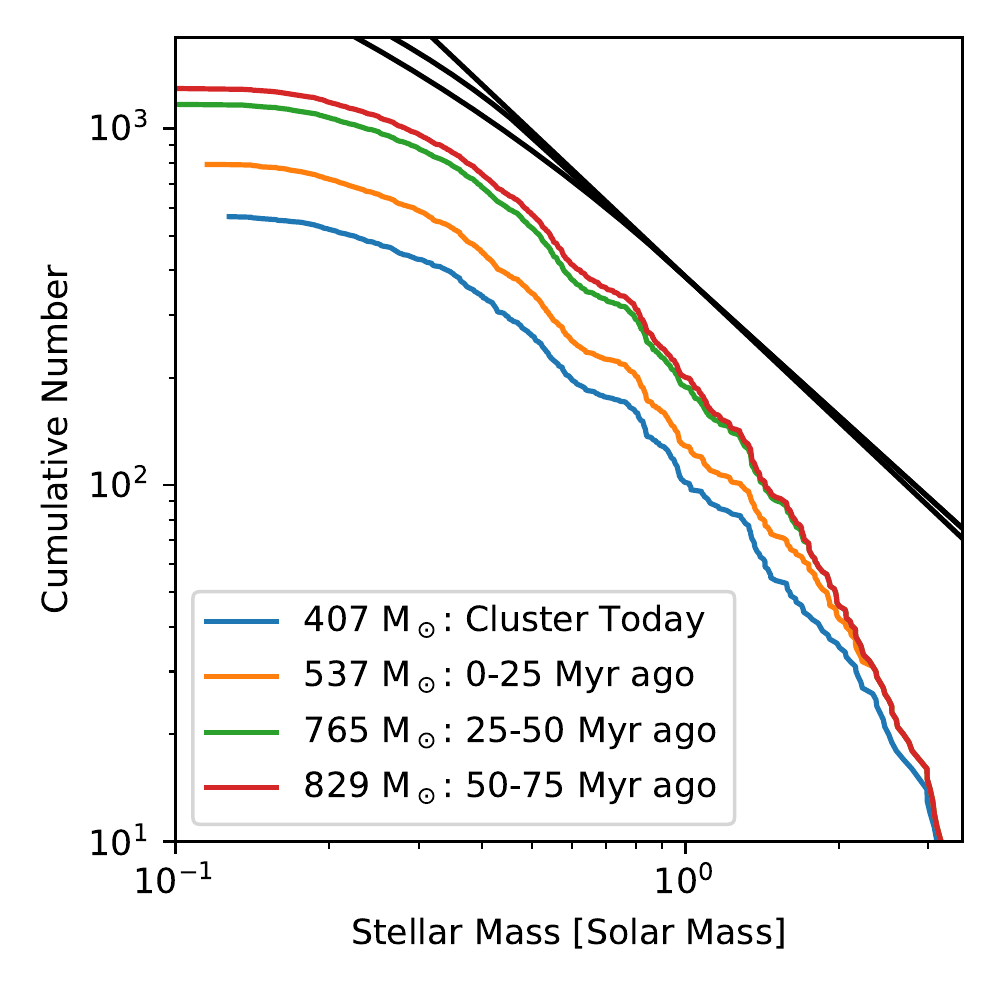}
    \caption{Cumulative mass function of the alpha Persei cluster today and in the past.  The black curves from top to bottom trace  \citet{1955ApJ...121..161S}, \citet{2001MNRAS.322..231K} and \citet{2003PASP..115..763C} initial-mass functions.}
    \label{fig:cum-mass-funk}
\end{figure}

Fig.~\ref{fig:cum-mass-funk} accumulates the mass lost from the cluster with the current mass function of the cluster to know how the mass function of the cluster has evolved.    Both the mass loss rate and the mean mass of the lost stars  have been increasing over the past 75~Myr.  From the distribution of masses in Fig.~\ref{fig:escape-mass}, we see that the mass distribution of the potential escapees before 75~Myr ago does not follow this trend.  From this and from the analysis in \S~\ref{sec:k-ages}, we conclude that the age of the cluster is a bit larger than 75~Myr.
The bulk of the stars lost have had masses much smaller than one solar mass.  Initially (75 Myr ago) the cluster had nearly 1,300 stars with a mean mass of 0.64~M$_\odot$.  Today the cluster has nearly 600 stars with a mean mass of 0.69~M$_\odot$.  The mean mass of the 710 lost stars is 0.59~M$_\odot$.  As we found in the Pleiades
\citep{pleiades}, 
 the cluster has a deficit of both low-mass and high-mass stars relative to three commonly used initial mass functions \citep{1955ApJ...121..161S,2001MNRAS.322..231K,2003PASP..115..763C}.

It is natural to compare the history of the alpha Persei cluster with that of the Pleaides (Paper I).  Both alpha Persei and the Pleiades started with about eight-hundred solar masses.   The Pleiades over the past 120~Myr has lost twenty percent of its initial mass,  while alpha Persei has lost half of its mass over just 80~Myr.  If we convert these to mass-loss timescales ($M/{\dot M})$, that of the Pleiades is about 200~Myr and that of alpha Persei is just 55~Myr.  Given that the two clusters started with similar masses and assuming a similar underlying stellar population, what could cause such a divergence during their evolution?  A hint is apparent in comparing Eq.~\ref{eq:1} and~\ref{eq:2} here with the values for the Pleiades in Eq.~1 and~2 of Paper I.  If we combine these measurements with a model of the local Galactic potential and the motion of the Sun with respect to the local standard of rest \citep{2008gady.book.....B}, we find that the maximum displacement of the alpha Persei cluster with respect to the Galactic plane is about 20~pc whereas that of the Pleiades is about 120~pc.  The alpha Persei cluster has spent a larger fraction of its history close to the Galactic plane at a smaller relative velocity whereas the Pleiades passed through the Galactic plane quickly about 36, 80 and 122~Myr ago.  These factors increase the frequency and intensity of the interactions with molecular clouds and sources of tidal forces accelerating the demise of this cluster relative to the Pleiades.  

\subsection{The Three Youngest Clusters}
\label{sec:young-clusters}

\begin{figure*}
\centering
    \centering
    \includegraphics[width=0.32\textwidth]{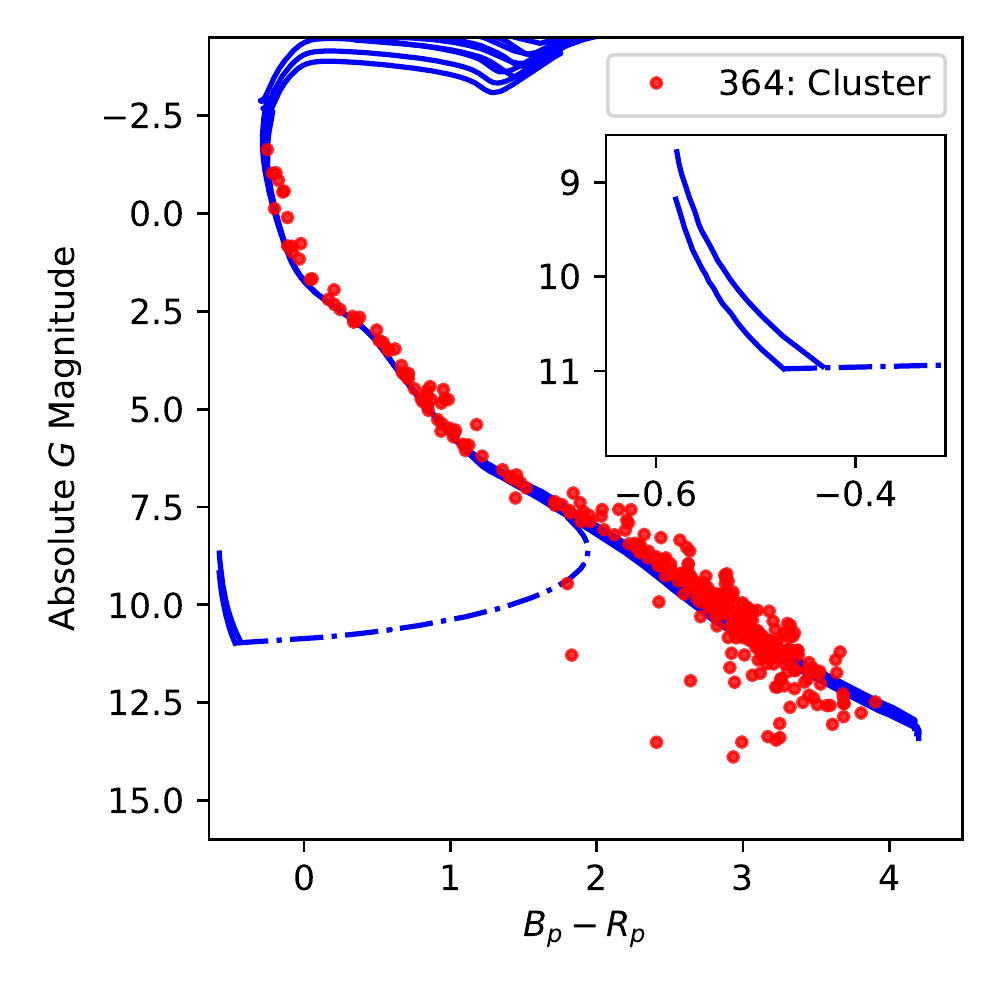}
    \includegraphics[width=0.32\textwidth]{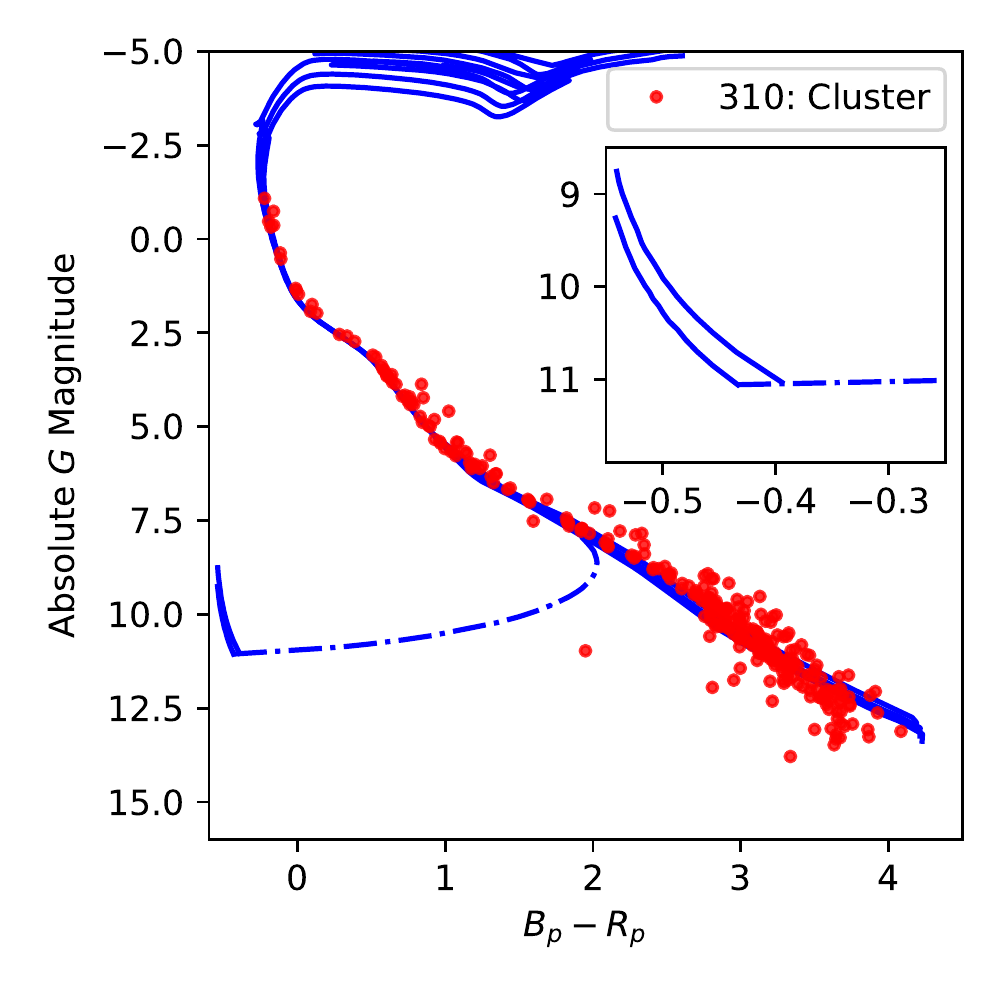}
    \includegraphics[width=0.32\textwidth]{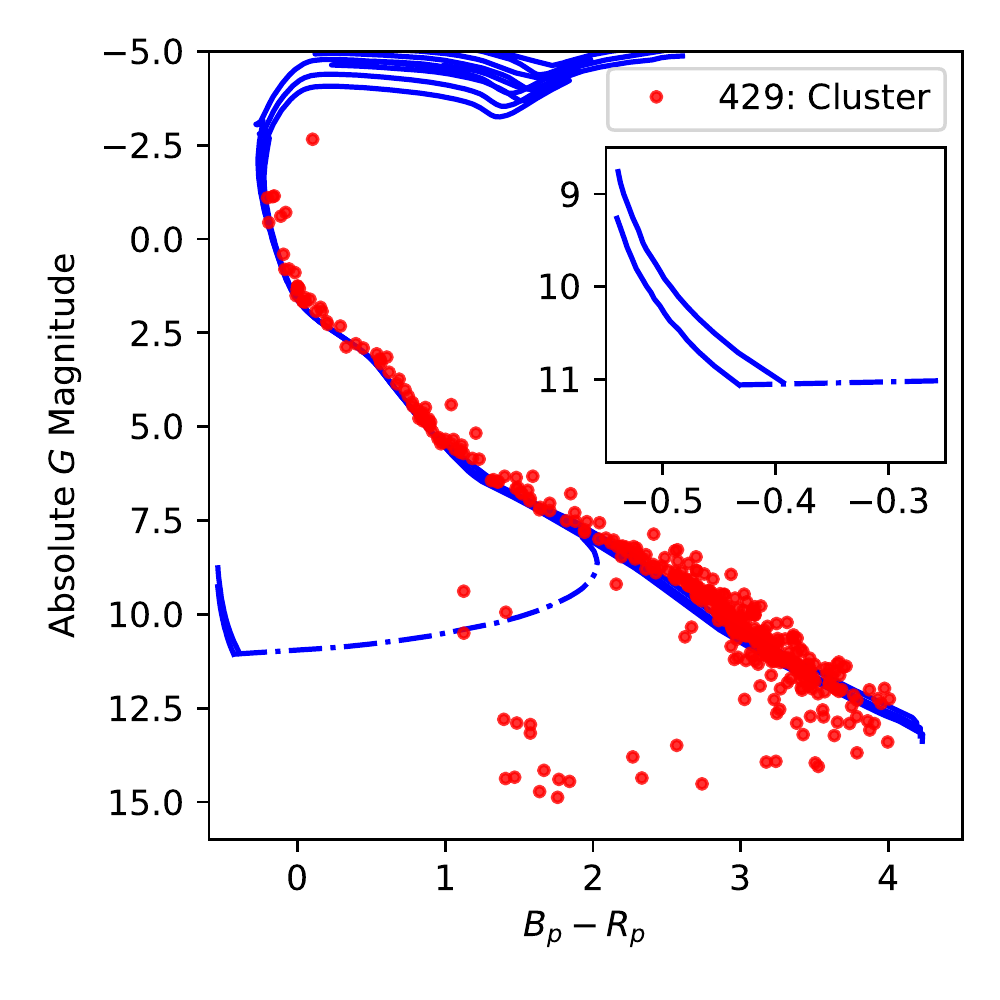}
    \caption{Colour-magnitude diagram of the young clusters (from left to right: NGC~2451A, IC~2391 and IC~2602) today using the volume-limited sample of those objects within 10~pc of the centre of the cluster and with proper motions lying within five milliarcseconds per year of the median.  
    The blue curves are Padova isochrones ($Z=0.0152$) with ages of 50, 60 and 70~Myr for NGC~2451A and with ages of 40, 50 and 60~Myr for IC~2391 and IC~2602 as well as Montreal white-dwarf models of 1.0 and 1.1 solar masses up to a cooling ages of 30~Myr (NGC) and 20~Myr (ICs).  
    No extinction has been applied to the models for NGC~2451A, an 
     extinction of $A_G=0.08$ and a reddening of $E(B_p-R_p)=0.04$ have been applied for IC~2391 and IC~2602.  To convert the apparent $G$ magnitudes to absolute, we use the median distance of the samples of 192~pc (NGC) and 152~pc (ICs) obtained from the parallaxes. A binary composed of the oldest white dwarf expected in the cluster and a main-sequence star is shown by the dot-dashed curves.}
    \label{fig:cmd-yc}
\end{figure*}

 We proceed to identify the stars in each cluster in a similar manner to what we did for alpha Persei.  First we identify all stars within one degree on the sky  and within five milliarcseconds per year in proper motion of the nominal position of each cluster in phase space (Tab.~\ref{tab:sample}).  We then find the median proper motion and distance of the stars within this sample, and define the cluster sample to contain all stars within 10~pc of the centre of the cluster and within five milliarcsesconds per year of the median cluster proper motion.  Fig.~\ref{fig:cmd-yc} depicts the colour-magnitude diagrams for each of these three youngest clusters.  The diagram shows that each of these three clusters is less rich than alpha Persei; therefore, the contrast of the escapees against the background in phase space is less pronounced, and it is impossible to find a threshold relative velocity where both the completeness and the purity of the sample exceed fifty percent, so to look at the colour-magnitude diagram of the escapee candidates we impose a relative velocity threshold of 2~km/s to construct our escapee candidate sample. Unfortunately, this choice does not allow us to estimate the total number of stars that have escaped the clusters, but it does give an impression of the escapee population and their relative rates of escape. 

\begin{figure*}
\centering
    \includegraphics[width=0.32\textwidth]{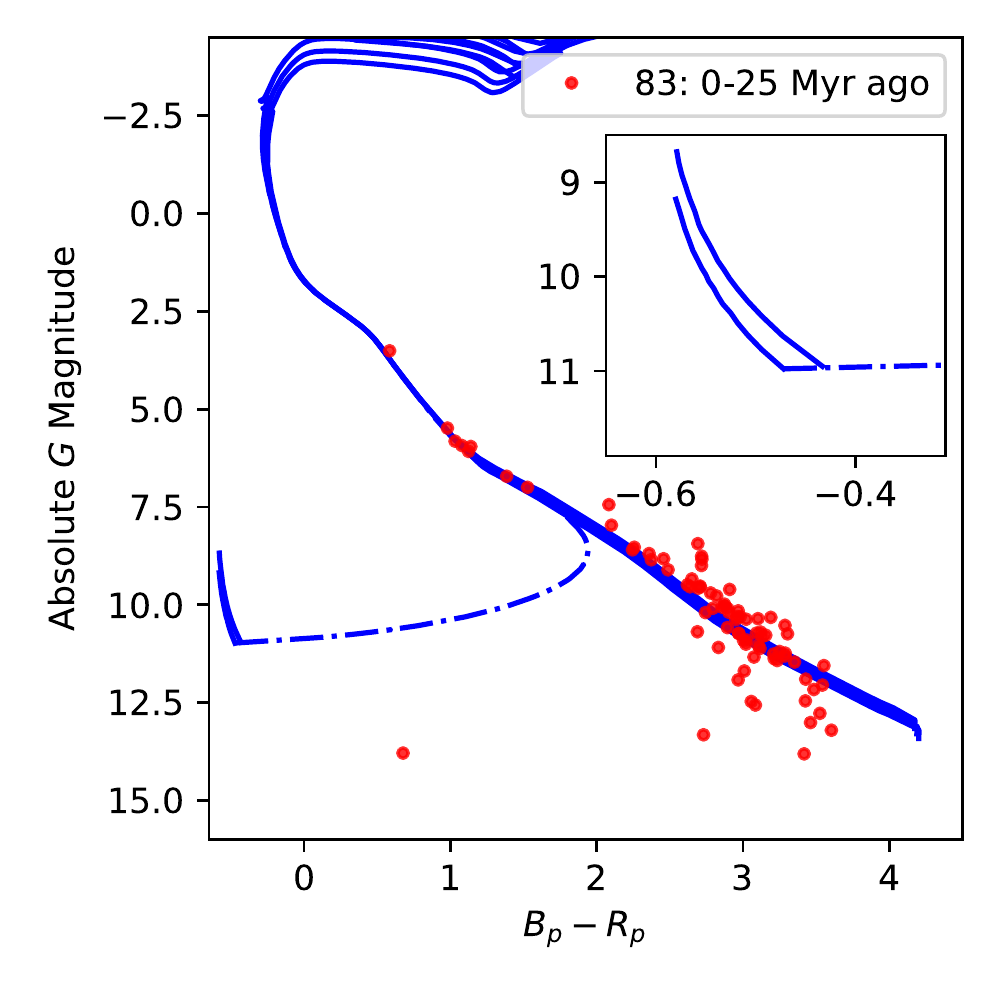}
    \includegraphics[width=0.32\textwidth]{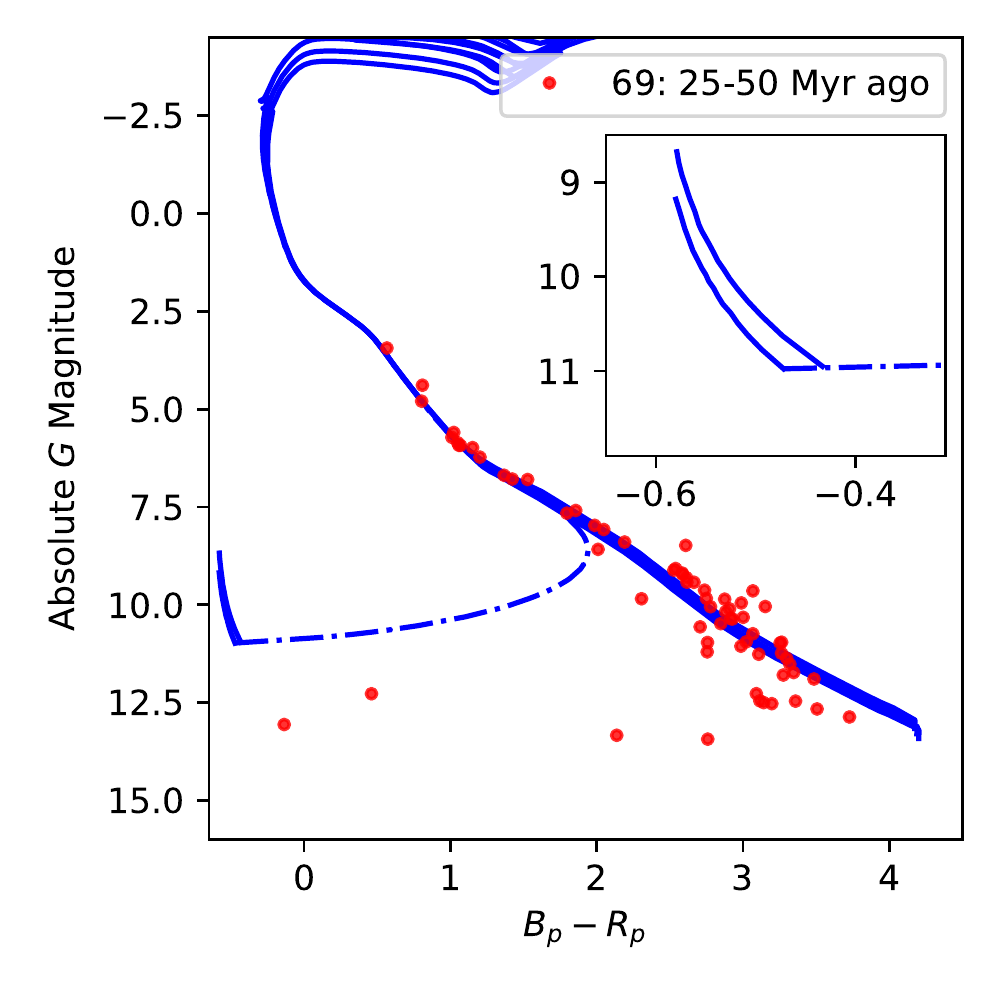}
    \includegraphics[width=0.32\textwidth]{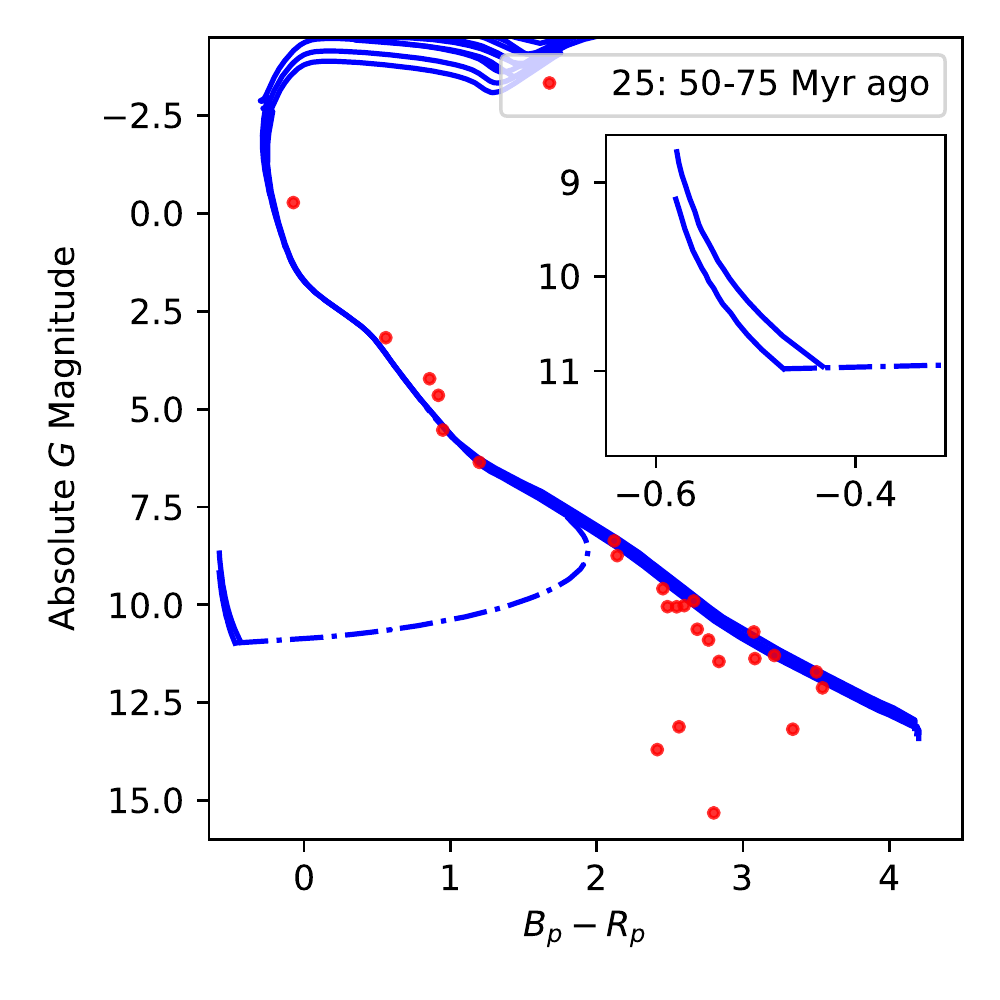}
    \includegraphics[width=0.32\textwidth]{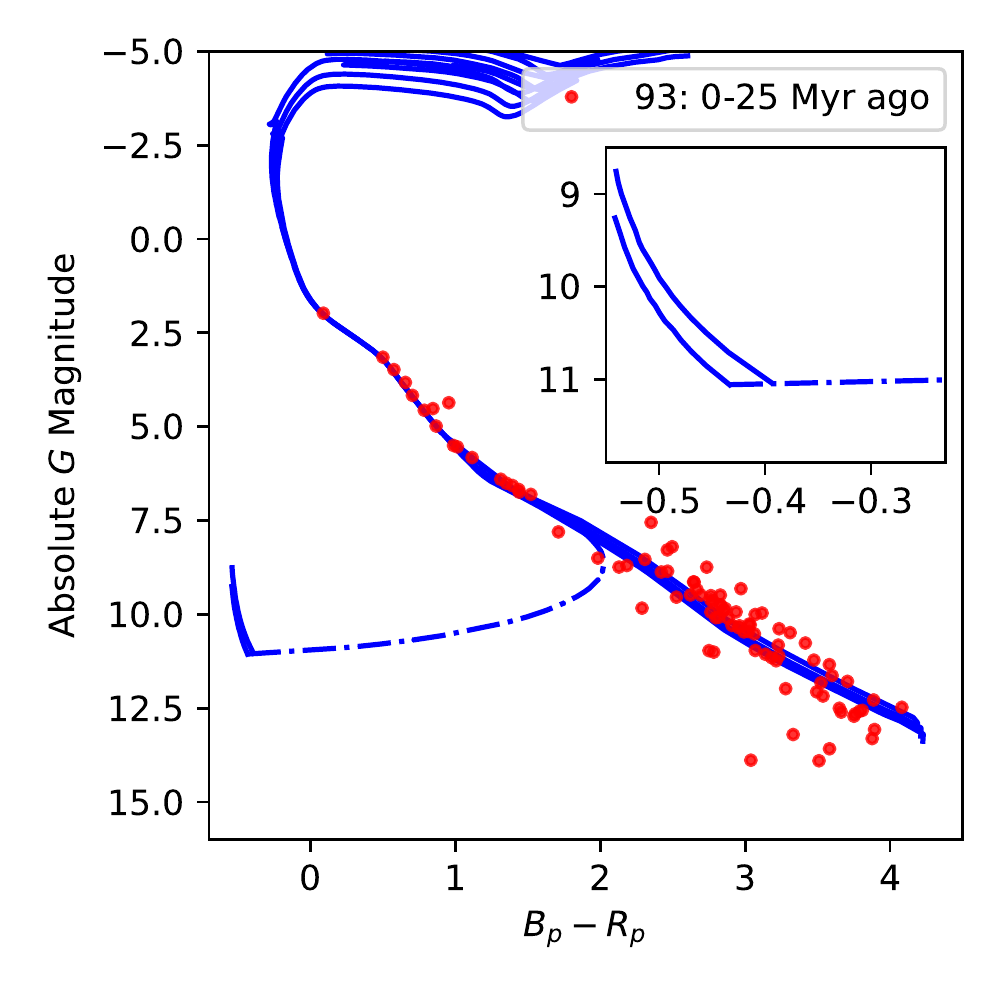}
    \includegraphics[width=0.32\textwidth]{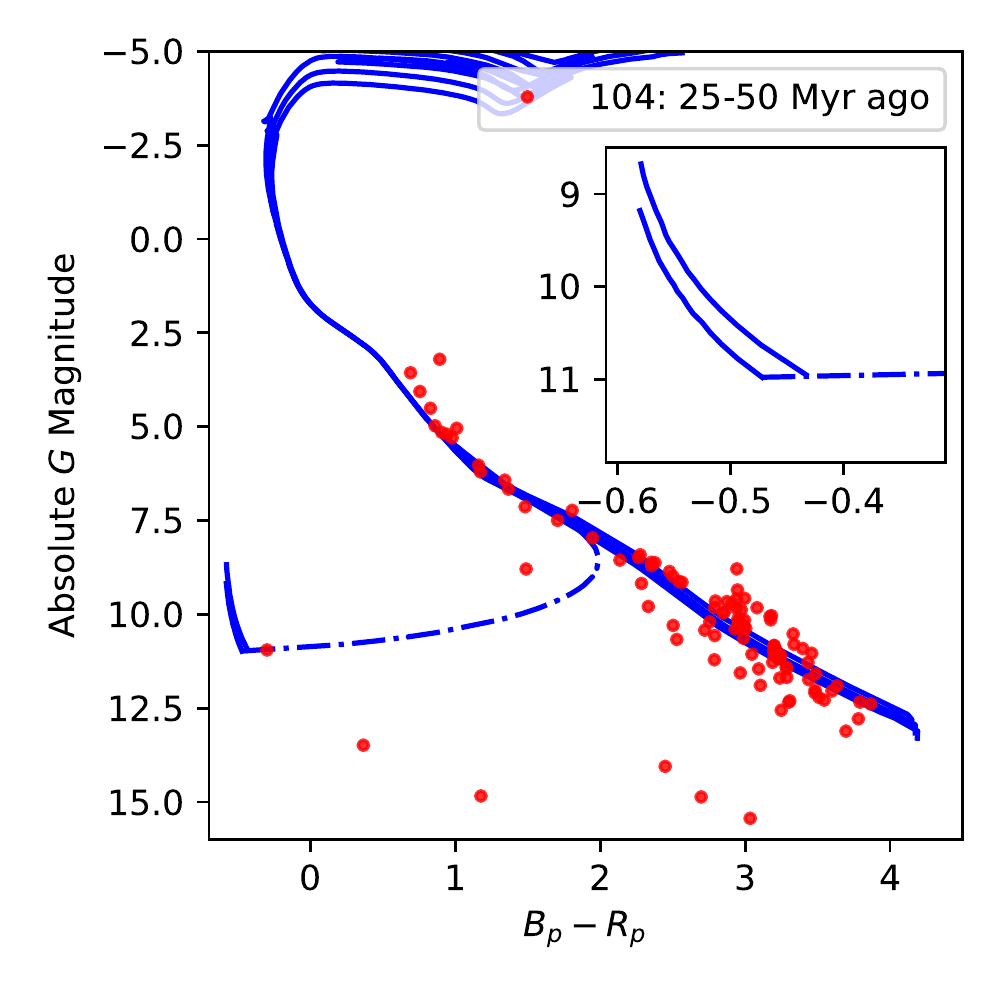}
    \includegraphics[width=0.32\textwidth]{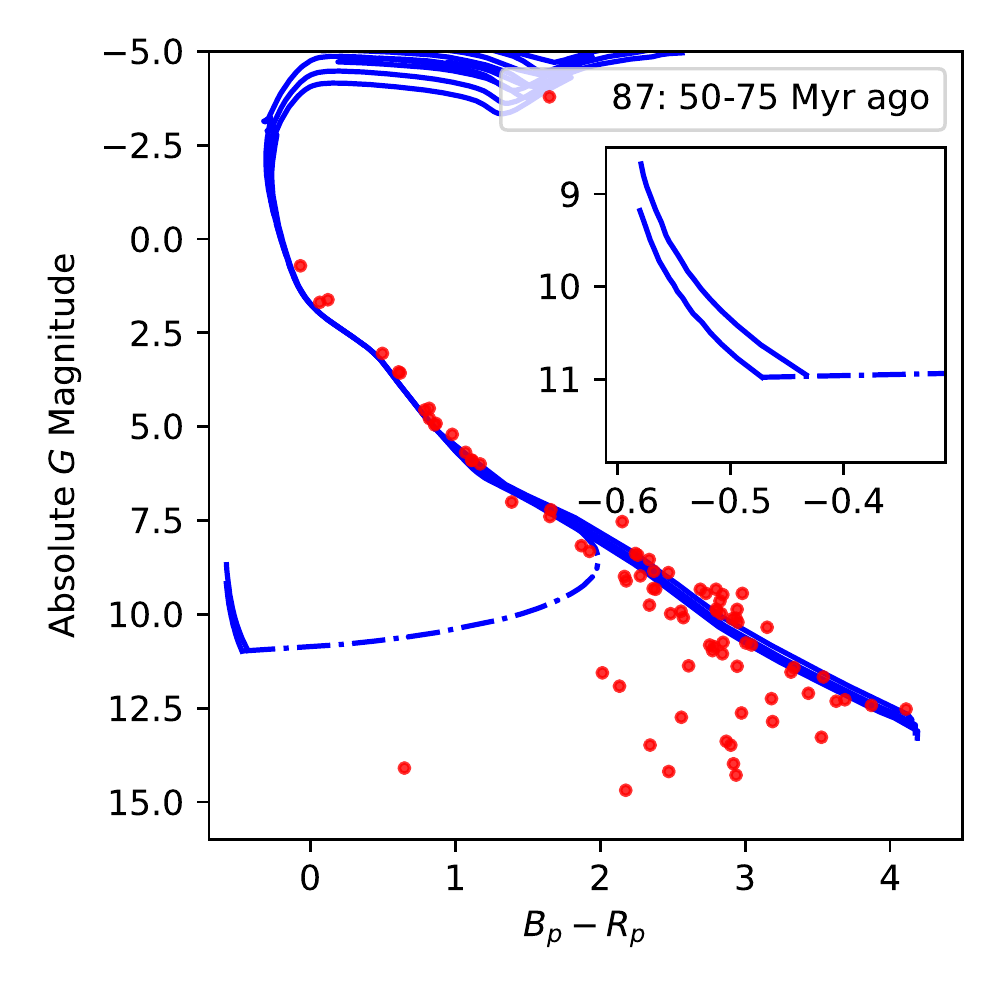}
    \includegraphics[width=0.32\textwidth]{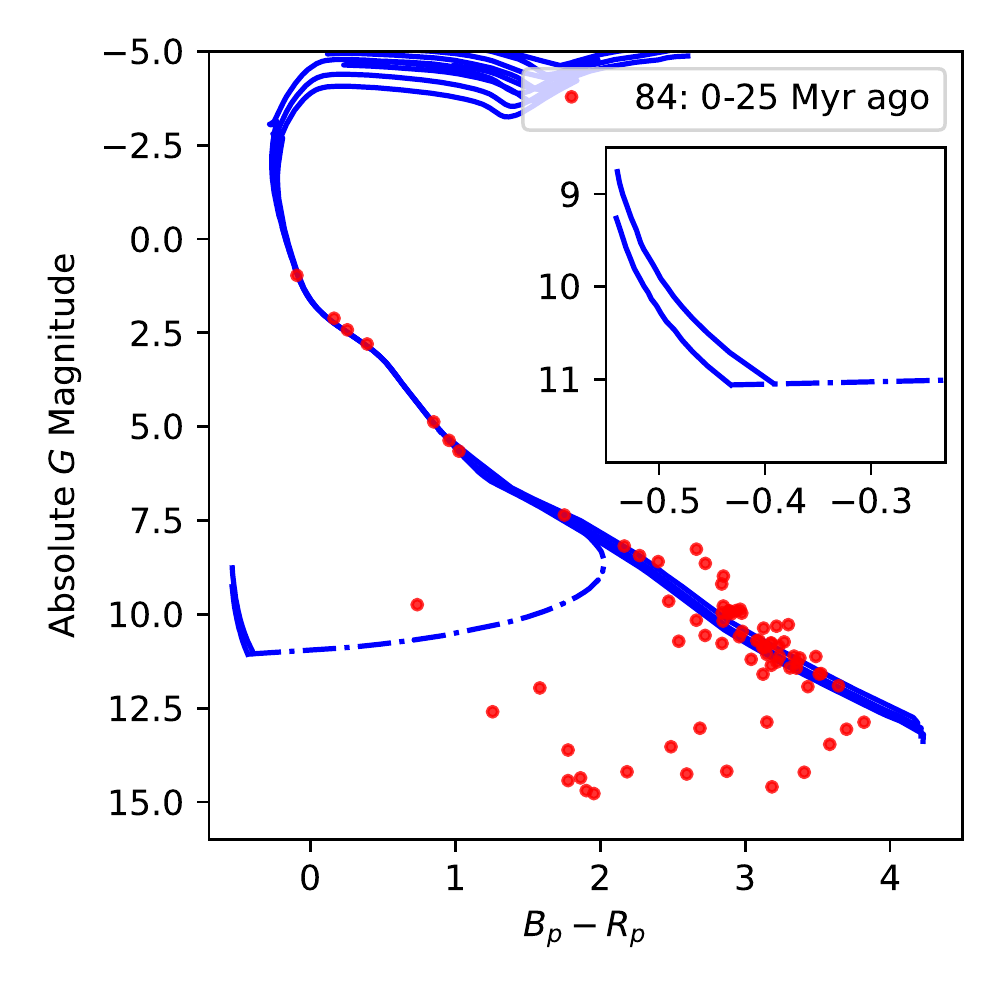}
    \includegraphics[width=0.32\textwidth]{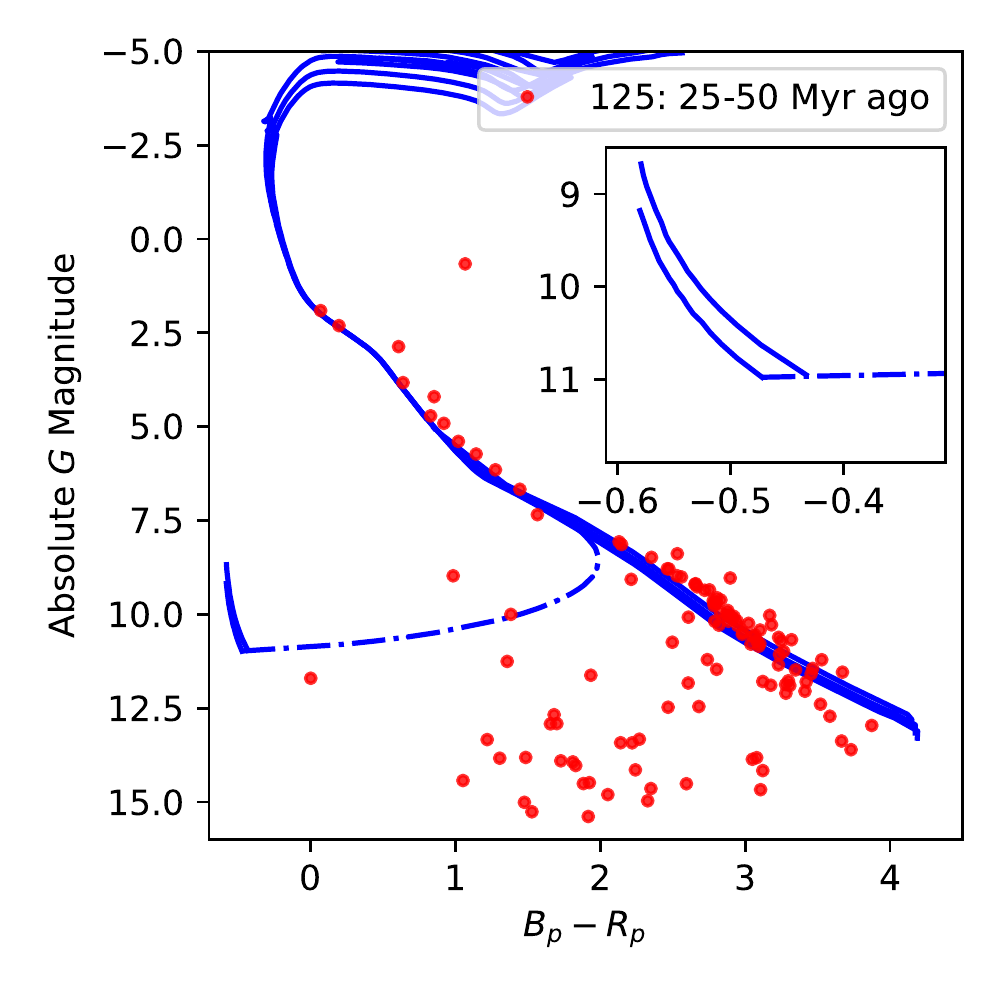}
    \includegraphics[width=0.32\textwidth]{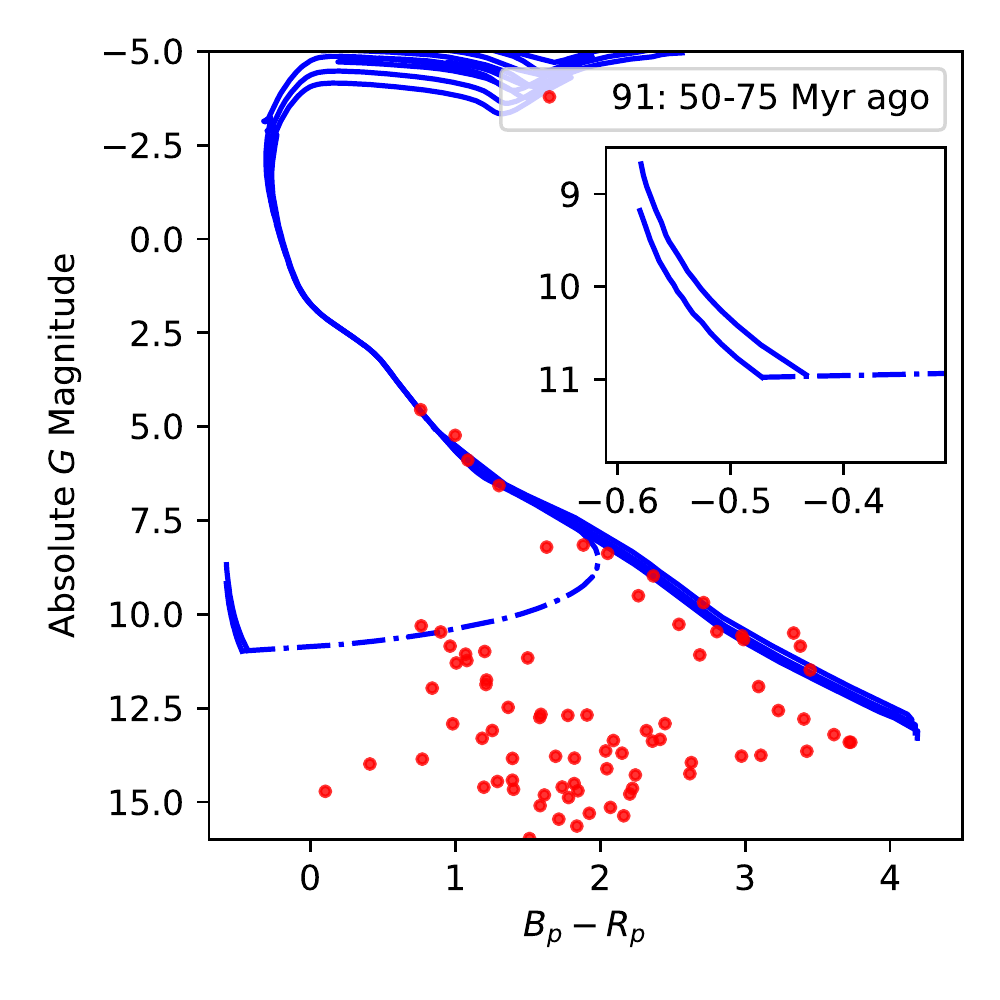}
    \caption{Colour-magnitude diagram of the escapees from the young clusters as a function of time from top to bottom: NGC~2451A, IC~2391 and IC~2602.  The blue curves are as in Fig.~\ref{fig:cmd-yc}.  The individual parallaxes of the objects are used to calculate the absolute magnitudes.}
    \label{fig:cmd-escape-yc}
\end{figure*}

From Fig.~\ref{fig:cmd-escape-yc} we see that for each cluster the population of escapee candidates that left the cluster up to 50~Myr ago is approximately constant with time and resembles the population within the cluster, especially at the low-mass end.   However, the candidate escapees that would have left earlier than 50~Myr ago do not strongly resemble the cluster population and rather look more like an older population or simply outliers.  These stars are simply on trajectories that can be traced back to intersect the current trajectory of the clusters before the clusters themselves existed.  The fact that these low-mass stars in the escapee candidate sample have already left the pre-main-sequence stage indicates that they are not actually cluster members.  The resemblance of the colour-magnitude diagram of the escapee candidates to that of the cluster especially at the low-mass end indicates a novel method to determine the ages of these clusters kinematically which we will now explore.

\section{Kinematic Ages}
\label{sec:k-ages}

\begin{figure}
    \centering
    \includegraphics[width=\columnwidth]{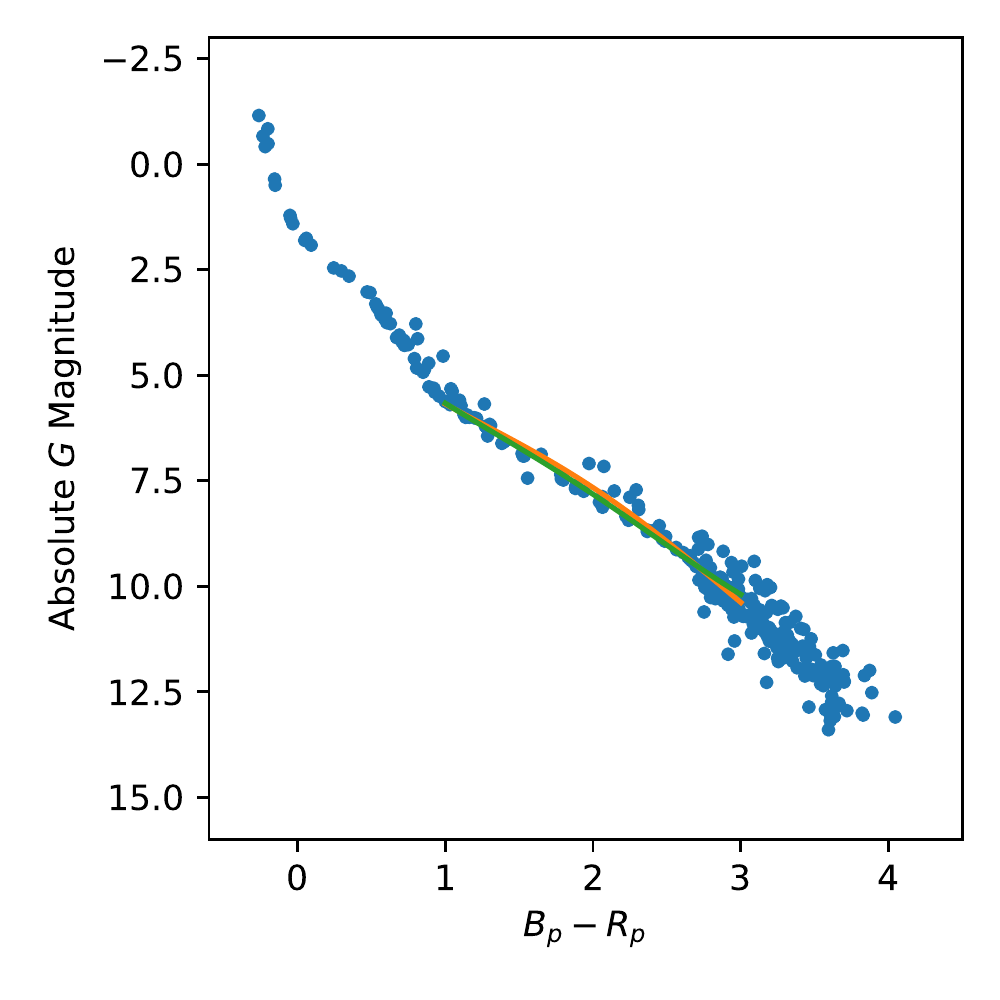}
    \caption{Fitting the lower main-sequence of the cluster IC~2391 using the traditional $\chi^2$ estimator in orange and the robust estimator $R^2$ in green.}
    \label{fig:cmd-fitter}
\end{figure}

Fig.~\ref{fig:cmd-escape}, Fig.~\ref{fig:cmd-escape-yc} as well as Fig.~9 from Paper I (for the Pleiades) show how the population of candidate escapees resembles the cluster population for escapees that left after the cluster formed.   To focus particularly on the lower-main sequence we look at stars in the cluster with dereddened colours $1<B_p-R_p<3$ and brighter than $G=6.3$ and $G=11.9$ at the blue and red end respectively.  We fit the lower main-sequence of the cluster stars with a quartic polynomial function of absolute magnitude against colour.  This function forms the fiducial main-sequence ridge line of the cluster.  Fig.~\ref{fig:cmd-fitter} shows the results of this fitting for the cluster IC~2391.  We fit the polynomial in two ways: the first minimizes the value of $\chi^2$ to find the best fit and the second uses a robust estimator that we denote $R^2$
\begin{equation}
    R^2 = n \left [ H\!L \left ( X_i \right ) \right ]^2 + n \left [ Q_n \left (X_i \right ) \right ]^2.
    \label{eq:chi2_robust}
\end{equation}
$R^2$ uses two alternative statistics based on pairs of measurements: the Hodges-Lehmann-Sen robust estimator of location $H\!L$ \citep{hodges1963,10.2307/2527532} and $Q_n$ robust  estimator of scale \citep{doi:10.1080/01621459.1993.10476408}, which are defined as 
\begin{eqnarray}
H\!L &=&\textrm{median}\left\{\frac{X_{i}+X_{j}}{2} :   i \leq j  \right\}, \\
Q_{n} &=& c_{n}\textrm{~first quartile of~}\left\{\left|X_{i}-X_{j}\right| :  i<j \right\},
\end{eqnarray}
where the value of $c_{n}$ depends on the size of the sample and scales the value of $Q_n$ to be the standard deviation in the case of normally distributed data. A python library is available at \url{https://github.com/UBC-Astrophysics/qn_stat} that uses fast algorithms (${\cal O}(n\log n)$) to determine both \citep{doi:10.1080/01621459.1993.10476408,10.1145/1271.319414}.   From the figure it is apparent that either statistic yields an acceptable fit to the lower main-sequence stars.

To estimate the age of the cluster, we use these two statistics to determine how the colour-magnitude diagram of the candidate escapees changes as the time of escape moves further into the past.  For the young clusters (NGC 2451A, IC 2391 and IC 2602) we look at stars that could have escaped the cluster in ranges of 10~Myr.  Fig.~\ref{fig:geom-age-pl} shows the values of $R^2$ and $\chi^2$ normalised by the number of escapees as a function of the maximum escape time of the population.  For NGC 2451A we see that if we allow candidate escapees that left the cluster more than 50~Myr ago into the sample, the colour-magnitude diagram begins to deviate significantly from that of the cluster using both the $R^2$ statistic in blue and the $\chi^2$ statistic in orange.  We take the time where both statistics increase dramatically and coincidentally as an estimate of the age of the cluster.  Although determining these kinematic ages from these statistics is somewhat subjective, the kinematic age does not depend on any assumptions of stellar evolution save the point that young low-mass stars take a while to reach the main sequence, so that stars from the cluster will appear brighter than typical stars of the same colour. 
\begin{figure*}
    \centering
    \includegraphics[width=0.32\textwidth]{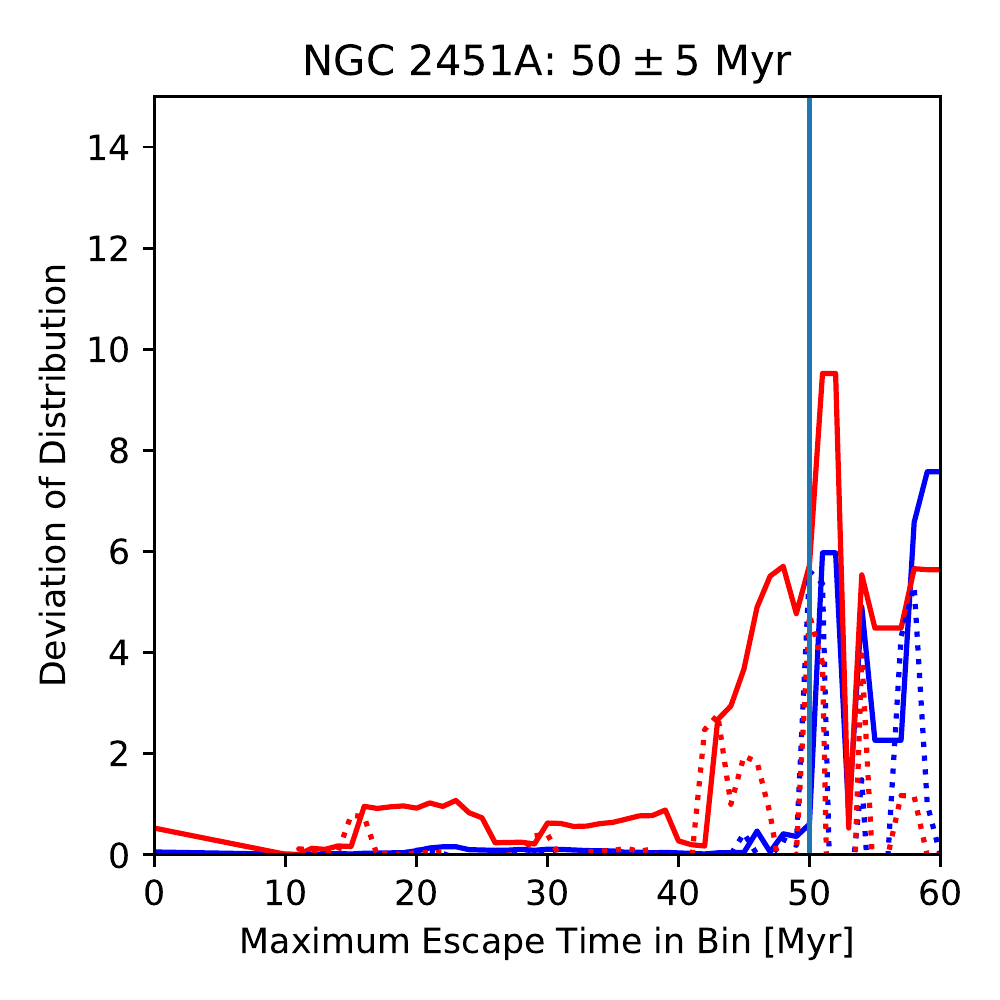}
    \includegraphics[width=0.32\textwidth]{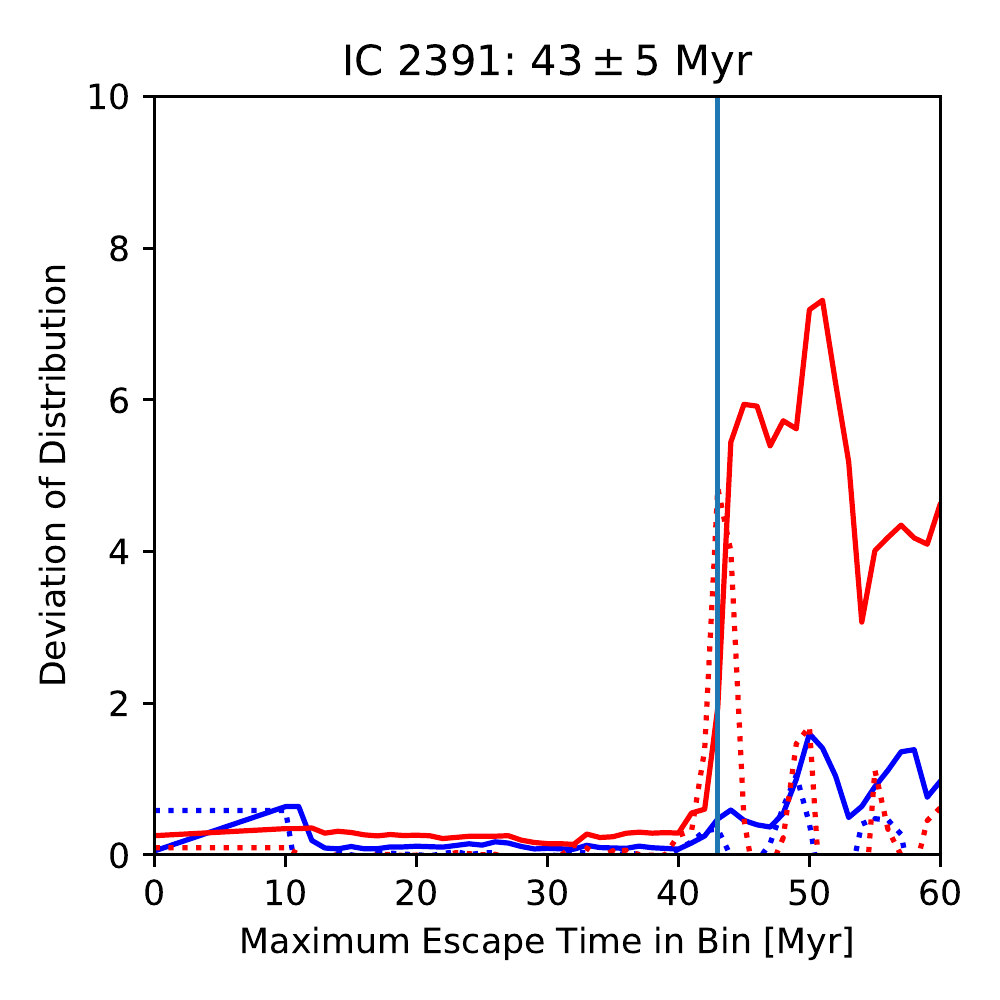}
    \includegraphics[width=0.32\textwidth]{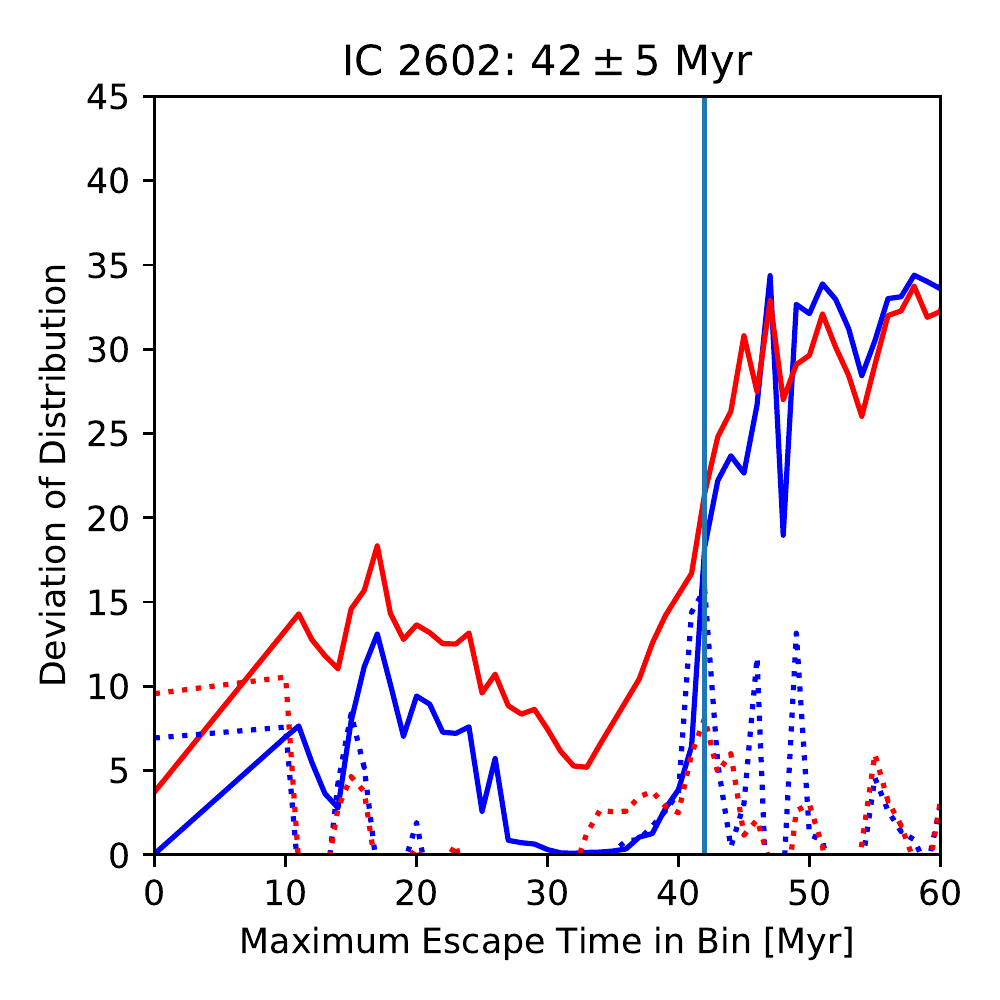}
    \caption{Kinematic-age determination for the youngest clusters studied.  The robust deviation statistic $R^2$ is depicted in blue solid lines, and the value of $\chi^2$ is shown in red solid lines.  The dotted lines show the value of Sobel-edge-detection filter that is used to estimate the kinematic age of each cluster.}
    \label{fig:geom-age}
\end{figure*}

For the older clusters alpha Persei and the Pleiades we constructed an all-sky catalogue of stars even further down the main sequence
\begin{verbatim}
SELECT * FROM gaiaedr3.gaia_source WHERE 
    (parallax > 3 AND parallax_over_error > 5 AND 
     bp_rp > 2 AND bp_rp < 3)
\end{verbatim}
because the more massive bluer stars in these clusters have already reached the main sequence, so they appear similar to the field stars.  We selected candidate escapees using the same parameters as for alpha Persei in this paper and the Pleiades from Paper I.  Because these clusters are richer than the younger clusters, we can use a more generous cut in three-dimensional velocity. Because of this and their age the escapees can wander further from the cluster since the cluster was born, so using a larger volume sample was warranted.  Otherwise, the procedure for the older clusters was the same as for the younger ones and the corresponding age determinations are depicted in Fig.~\ref{fig:geom-age-pl}.  To determine the uncertainties in these kinematic age determinations, we perform one hundred bootstrap resamplings of the escapee distribution from each of the clusters and measure the kinematic age of the cluster by finding the peaks in the Sobel-edge-detection value (shown as dotted curves in the figures) as a function of time for each cluster.  The standard deviation of the resulting distributions is typically about 5~Myr for the young clusters and slightly larger for alpha Persei and the Pleiades, which we take as uncertainties in this techinique.
\begin{figure*}
    \centering
    \includegraphics[width=0.48\textwidth]{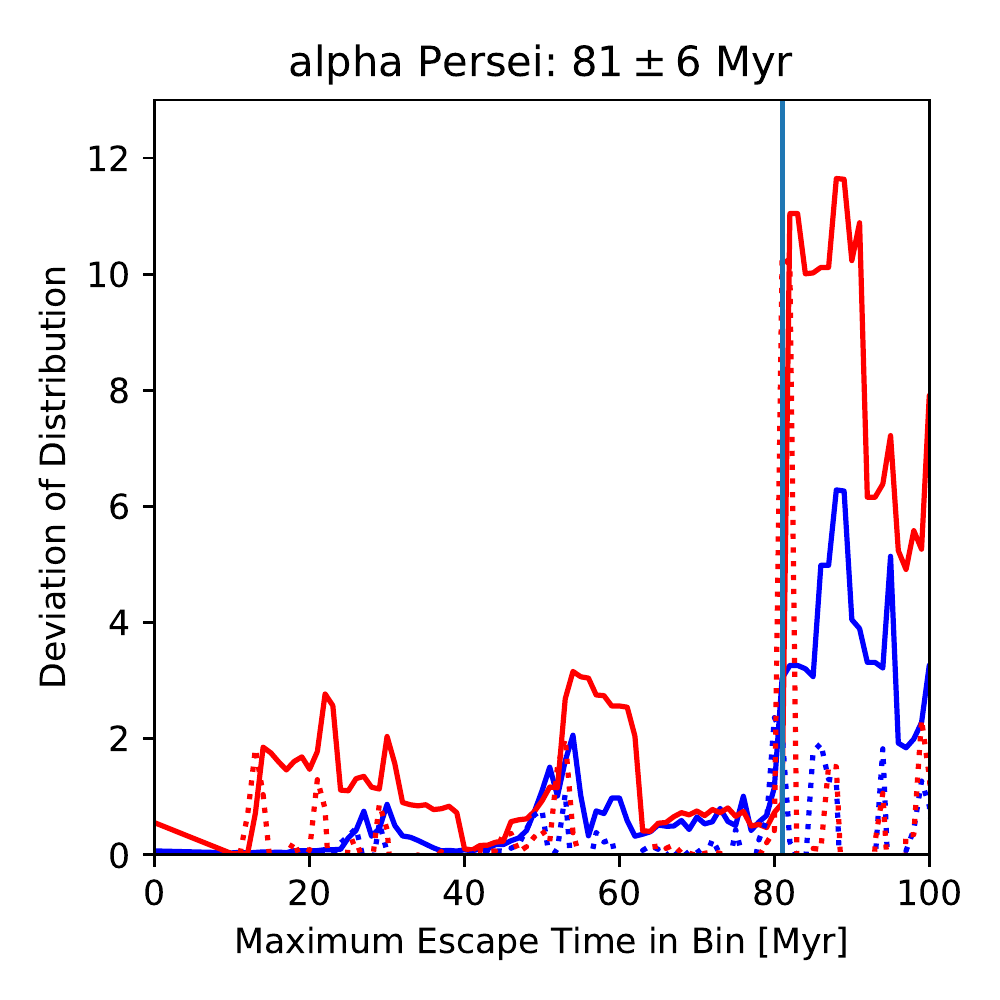}
    \includegraphics[width=0.48\textwidth]{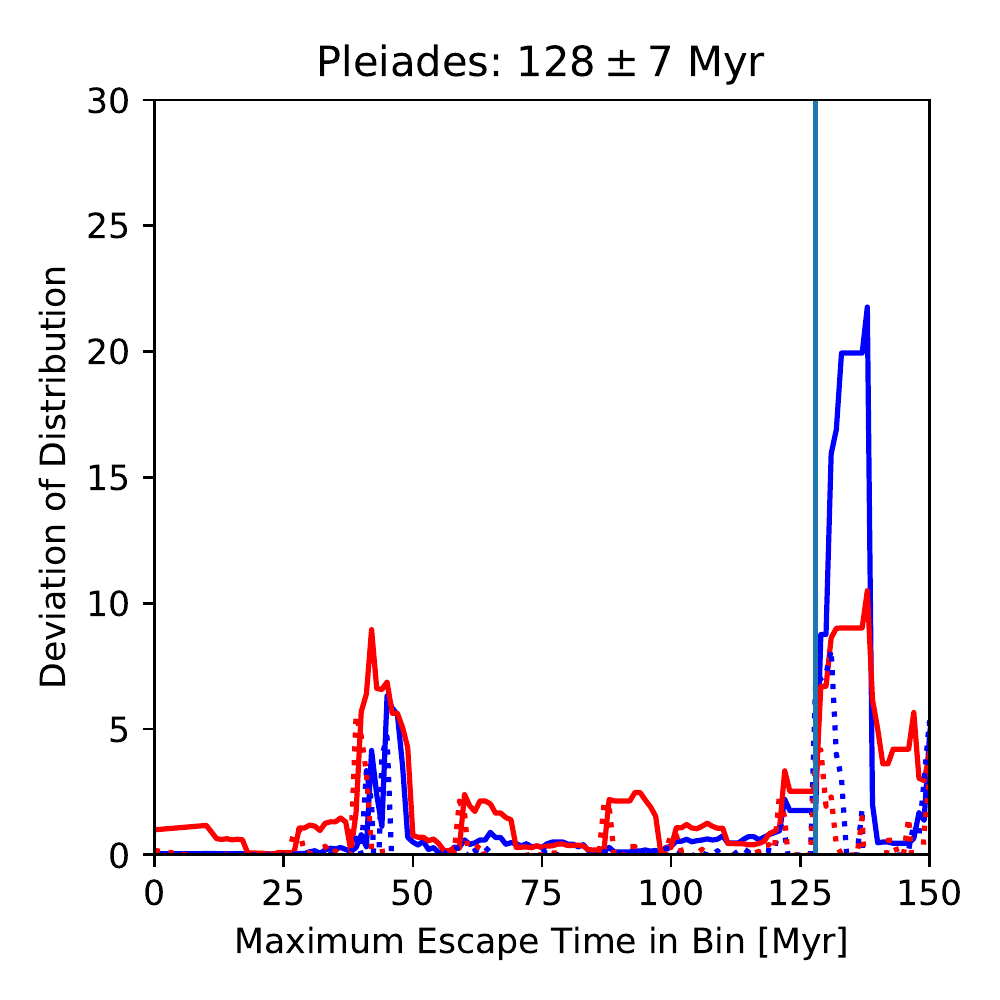}
    \caption{Kinematic-age determination for the alpha Persei and Pleiades clusters}
    \label{fig:geom-age-pl}
\end{figure*}

We can compare these kinematic ages with the properties of the orbits of these clusters above and below the Galactic plane.  As we discussed earlier the Pleiades reaches a height of 120~pc above the plane and passed through the plane about 36, 80 and 122~Myr ago.  The earliest passage coincides approximately with the kinematic age of the cluster of 128~Myr.  We find a similar coincidence for the clusters NGC~2451A and IC~2602 which last passed through the plane 47 and 41~Myr ago respectively with kinematic ages of 50 and 42~Myr.  On the other hand, for the clusters alpha Persei  (with passages 27, 69 and 112~Myr ago) and IC~2391 (with passages 33 and 76~Myr ago), the cluster passages do not coincide with the age estimates of 81 and 43~Myr respectively.  The difference is that NGC~2451A and IC~2602 reach a large distance (80 and 95~pc respectively) from the plane like the Pleiades, so their formation should coincide closely with passage through the plane.  In fact they coincide within a few~Myr. On the other hand alpha Persei and IC~2391 travel only up to 20 and 30~pc from the plane, so they could have formed at any point along their orbit.

\section{Conclusions}
\label{sec:conclusions}

We have built upon our previous analysis of the Pleiades cluster in Paper I to examine the other four young clusters (younger than 200~Myr) within 200~pc of the Sun.  Only the cluster alpha Persei is about as rich as the Pleiades, so we could perform a similar analysis to obtain an estimate of all of the stars lost from this cluster as we did for the Pleiades.  We found that the mass-loss rate from the alpha Persei cluster is about four times larger than the Pleiades.  It has already lost more than half of its original mass.   We also found that the orbit of this cluster does not bring it nearly as far from the Galactic plane as the Pleiades travels.  Its proximity to the plane may account for its rapid mass loss.  Among the escapees from the alpha Persei cluster, we found two candidate massive white dwarfs \citep{alphaperWD}.  Because the younger clusters are less rich, we chose more conservative criteria to determine the sample, limiting the number of falsely identified escapees at the expense of missing many candidates.  For the younger clusters we found no massive white dwarf candidates, but we did find several potential white-dwarf-main-sequence binaries for future follow-up.

An examination of the colour-magnitude diagram of the escapees shows that the escapees not surprisingly occupy a similar part of the diagram especially for the low-mass stars which have not yet settled on the main sequence in these young clusters distinguishing them from the older field stars.  We build a novel method to determine the ages of this clusters upon this observation.  We develop two estimators to quantify how much the colour-magnitude diagrams of two samples resemble each other and compare escapees over a narrow range of escape times and colours with cluster stars to determine for how long our escapee sample resembles the cluster.  The timescale over which the two samples resemble each other is the lifetime of the cluster.   In general we find good agreement between ages determined with isochrones and these kinematic ages; however, the new technique yields  uncertainties three to four times smaller.   We find for the clusters that travel far from the Galactic plane (the Pleiades, NGC~2451A and IC~2602), the formation of the cluster coincides with when the cluster was in the Galactic plane within a few Myr, supporting these age determinations.

We have only scratched the surface with these escapee catalogues.  They can be tested and analysed further by correlating them with other spectroscopic and photometric catalogues.  As we discussed in Paper I, this 5D technique may be limited to just these clusters, because of measurement uncertainties which we have neglected here.   Perhaps, we could make further progress by including information from the colours and magnitudes of the stars to inform the definition of the escapee sample and also including the effects of uncertainties, to assess an {\it a priori} probability of cluster membership from the astrometry alone and an {\it a posteriori} probability including the intrinsic properties of the stars.  We have seen how a simple dynamical model of the Solar Neighbourhood can help build deeper conclusions about the history of these clusters.  Such a model could be included in the identification of escapees as well. A key advantage of the simple method presented here is that it is analytic and not iterative, so in principle, one can search the entire Gaia EDR3 catalogue for escapees from a particular cluster quickly.   It may be possible to identify clusters in action-angle coordinates and look for candidate escapees in this projection, building dynamics into this kinematic picture.  

\section*{Acknowledgements}

This work was supported in part by NSERC Canada and Compute Canada.

This research has made use of the SIMBAD and Vizier databases, operated at CDS, Strasbourg, France and the Montreal White Dwarf Database produced and maintained by Prof. Patrick Dufour (Universit\'e de Montr\`eal) and Dr. Simon Blouin (LANL), 


This work has made use of data from the European Space Agency (ESA) mission
{\it Gaia} (\url{https://www.cosmos.esa.int/gaia}), processed by the {\it Gaia}
Data Processing and Analysis Consortium (DPAC,
\url{https://www.cosmos.esa.int/web/gaia/dpac/consortium}). Funding for the DPAC
has been provided by national institutions, in particular the institutions
participating in the {\it Gaia} Multilateral Agreement.

\section*{Data Availability}

The data used in this study was obtained from the ESA Gaia Archive using the commands outlined in the appendix and processed using TOPCAT.  The resulting catalogues and derived quantities are included in the catalogues.



\bibliographystyle{mnras}
\bibliography{main,parsec,yulwd} 



\onecolumn
\appendix

\section{Creating the Gaia EDR3 young cluster sample}

The samples were generated using the following ADQL queries to the ESA Gaia archive.\\
\\
IC 2391: 
\begin{verbatim}
SELECT * FROM gaiaedr3.gaia_source WHERE 
    (parallax > 4.312 AND parallax_over_error >5 AND 1=CONTAINS(POINT('ICRS', ra, dec),
                 CIRCLE('ICRS', 130.133, -53.033, 23.0 )))  OR 
    (parallax> 5.165 AND parallax_over_error > 5 AND 1=CONTAINS(POINT('ICRS', ra, dec),
                 CIRCLE('ICRS', 130.133, -53.033, 52.2  )))
\end{verbatim}
IC 2602: 
\begin{verbatim}
SELECT * FROM gaiaedr3.gaia_source WHERE 
    (parallax > 4.304 AND parallax_over_error >5 AND 1=CONTAINS(POINT('ICRS', ra, dec),
                 CIRCLE('ICRS', 160.742, -64.400, 22.9 )))  OR 
    (parallax> 5.157 AND parallax_over_error > 5 AND 1=CONTAINS(POINT('ICRS', ra, dec),
                 CIRCLE('ICRS', 160.742, -64.400, 52.0  )))
\end{verbatim}
alpha Persei: 
\begin{verbatim}
SELECT * FROM gaiaedr3.gaia_source WHERE 
    (parallax > 3.923 AND parallax_over_error >5 AND 1=CONTAINS(POINT('ICRS', ra, dec),
                 CIRCLE('ICRS',  51.675,  48.800, 21.2 )))  OR 
    (parallax> 4.714 AND parallax_over_error > 5 AND 1=CONTAINS(POINT('ICRS', ra, dec),
                 CIRCLE('ICRS',  51.675,  48.800, 43.3  )))
\end{verbatim}
NGC 2451A: 
\begin{verbatim}
SELECT * FROM gaiaedr3.gaia_source WHERE 
    (parallax > 3.657 AND parallax_over_error >5 AND 1=CONTAINS(POINT('ICRS', ra, dec),
                 CIRCLE('ICRS', 115.800, -38.400, 19.8 )))  OR 
    (parallax> 4.393 AND parallax_over_error > 5 AND 1=CONTAINS(POINT('ICRS', ra, dec),
                 CIRCLE('ICRS', 115.800, -38.400, 38.3  )))
\end{verbatim}

After obtaining each sample from the Gaia archive, the following commands in TOPCAT \citep{2005ASPC..347...29T} create the additional vectorial quantities used in the paper, the velocity ${\bf v}_{3D}$:
\begin{verbatim}
uvw icrsToGal(astromUVW(array(ra, dec, parallax, pmra, pmdec, dr2_radial_velocity)))
\end{verbatim}
the velocity of the star in the plane of the sky, ${\bf v}_{2D}$,
\begin{verbatim}
uvw0 icrsToGal(astromUVW(array(ra, dec, parallax, pmra, pmdec, 0)))
\end{verbatim}
and the position, ${\bf r}$
\begin{verbatim}
xyz icrsToGal(astromXYZ(ra, dec, parallax))
\end{verbatim}
All of these quantities have been converted to Galactic coordinates.

\section{Description of Cluster and Escapee Catalogues}

The catalogues contain all of the Gaia EDR3 fields and the following additional fields:
\medskip

\noindent Vectorial quantities in Galactic coordinates
\begin{itemize}
    \item[uvw] ${\bf v}_{3D}$ defined as above in km/s.
    \item[uvw0] ${\bf v}_{2D}$ defined as above in km/s.
    \item[xyz] ${\bf r}$ defined as above in parsecs.
    \item[deltav] ${\bf \Delta v} = {\bf v}_{2D}-{\bf v}_{\textrm{cluster}}$ in km/s.
    \item[deltav3d] $\Delta \hat {\bf v}_\textrm{3D} =  {\bf \Delta v}  + v_r \hat {\bf r}$ as defined in Eq.~\ref{eq:9} in km/s
    \item[deltar] $\Delta {\bf r} = {\bf r}-{\bf r}_\textrm{cluster}$ in parsecs
    \end{itemize}
Scalar quantities
    \begin{itemize}
    \item[tmin] $t_\textrm{min}$ as defined in Eq.~\ref{eq:5} in Myr
    \item[vr] $v_r$ as defined in Eq.~\ref{eq:7} in km/s
    \item[dmin] $d_\textrm{min}$ as calculated from Eq.~\ref{eq:4} in parsecs.
    \end{itemize}
Boolean quantities
    \begin{itemize}
    \item[current] Boolean value to indicate current cluster members
\end{itemize}

\bsp	
\label{lastpage}

\end{document}